\documentclass{aa}
\usepackage{graphicx}
\begin{document}
\title{Accurate photometric redshifts for the CFHT Legacy Survey calibrated
  using the VIMOS VLT Deep Survey \thanks{Based on data obtained with
  the European Southern Observatory on Paranal, Chile, and on
  observations obtained with {\sc MegaPrime/Megacam}, a joint project
  of CFHT and CEA/DAPNIA, at the Canada-France-Hawaii Telescope (CFHT)
  which is operated by the National Research Council (NRC) of Canada,
  the Institut National des Science de l'Univers of the Centre
  National de la Recherche Scientifique (CNRS) of France, and the
  University of Hawaii. This work is based in part on data products
  produced at {\sc TERAPIX} and the Canadian Astronomy Data Centre as
  part of the Canada-France-Hawaii Telescope Legacy Survey, a
  collaborative project of NRC and CNRS. }}

\author{
O. Ilbert \inst{1,2}
\and S. Arnouts \inst{2}
\and H.J. McCracken \inst{3,4}
\and M. Bolzonella  \inst{1} 
\and E. Bertin  \inst{3,4}
\and O. Le F\`evre \inst{2}
\and Y. Mellier \inst{3,4}
\and G. Zamorani \inst{5}
\and R. Pell\`o \inst{6}
\and A. Iovino \inst{7}
\and L. Tresse \inst{2}
\and D. Bottini \inst{8}
\and B. Garilli \inst{8}
\and V. Le Brun \inst{2}
\and D. Maccagni \inst{8}
\and J.P. Picat \inst{6}
\and R. Scaramella \inst{9}
\and M. Scodeggio \inst{8}
\and G. Vettolani \inst{10}
\and A. Zanichelli \inst{10}
\and C. Adami \inst{2}
\and S. Bardelli  \inst{5}
\and A. Cappi    \inst{5}
\and S. Charlot \inst{11,3}
\and P. Ciliegi    \inst{5}  
\and T. Contini \inst{6}
\and O. Cucciati \inst{7}
\and S. Foucaud \inst{8}
\and P. Franzetti \inst{8}
\and I. Gavignaud \inst{11}
\and L. Guzzo \inst{8}
\and B. Marano     \inst{1}  
\and C. Marinoni \inst{2,12}
\and A. Mazure \inst{2}
\and B. Meneux \inst{2}
\and R. Merighi   \inst{5} 
\and S. Paltani \inst{13,14}
\and A. Pollo \inst{2}
\and L. Pozzetti    \inst{5} 
\and M. Radovich \inst{15}
\and E. Zucca    \inst{5}
\and M. Bondi \inst{10}
\and A. Bongiorno \inst{1}
\and G. Busarello \inst{15}
\and S. De La Torre \inst{2}
\and L. Gregorini \inst{7}
\and F. Lamareille \inst{6}
\and G. Mathez \inst{6}
\and P. Merluzzi \inst{15}
\and V. Ripepi \inst{15}
\and D. Rizzo \inst{6}
\and D. Vergani \inst{8}
}

\offprints{O.~Ilbert, e-mail: olivier.ilbert1@bo.astro.it}
   
\institute{
Universit\`a di Bologna, Dipartimento di Astronomia - Via Ranzani 1, 40127, Bologna, Italy
\and
Laboratoire d'Astrophysique de Marseille, UMR 6110 CNRS-Universit\'e de Provence, BP8, 13376 Marseille Cedex 12, France
\and
Institut d'Astrophysique de Paris, UMR7095 CNRS, Universit\'e Pierre \&
Marie Curie, 98 bis boulevard Arago, 75014 Paris, France
\and
Observatoire de Paris, LERMA, 61 Avenue de l'Observatoire, 75014 Paris, France
\and
INAF-Osservatorio Astronomico di Bologna - Via Ranzani 1, 40127, Bologna, Italy
\and
Laboratoire d'Astrophysique de l'Observatoire Midi-Pyr\'en\'ees, UMR 5572, 14 avenue E. Belin, 31400 Toulouse, France
\and
INAF-Osservatorio Astronomico di Brera - Via Brera 28, Milan, Italy
\and
IASF-INAF - via Bassini 15, 20133, Milano, Italy
\and
INAF-Osservatorio Astronomico di Roma - Via di Frascati 33, 00040, Monte Porzio Catone, Italy
\and
IRA-INAF - Via Gobetti,101, 40129, Bologna, Italy
\and
European Southern Observatory, Garching, Germany
\and
Centre de Physique Th\'eorique, Marseille, France
\and
Integral Science Data Centre, ch. d'\'Ecogia 16, CH-1290 Versoix
\and
Geneva Observatory, ch. des Maillettes 51, CH-1290 Sauverny
\and
INAF-Osservatorio Astronomico di Capodimonte - Via Moiariello 16, 80131, Napoli, Italy
}

\date{Received ... / Accepted ... }
\titlerunning{}
\authorrunning{Ilbert et al.}

\abstract{We present and release photometric redshifts for an uniquely
  large and deep sample of 522286 objects with $i'_{AB}\le 25$ in the
  Canada-France Legacy Survey ``Deep Survey'' fields D1, D2, D3, and
  D4, which cover a total effective area of 3.2 $\deg^2$. We use 3241
  spectroscopic redshifts with $0 \leq z \leq 5$ from the VIMOS VLT
  Deep Survey as a calibration and training set to derive these
  photometric redshifts. Using the ``Le Phare'' photometric redshift
  code, we devise a robust calibration method based on an iterative
  zero-point refinement combined with a template optimisation
  procedure and the application of a Bayesian approach. This method
  removes systematic trends in the photometric redshifts and
  significantly reduces the fraction of catastrophic errors (by a
  factor of 2.3), a significant improvement over traditional
  methods. We use our unique spectroscopic sample to present a
  detailed assessment of the robustness of the photometric redshift
  sample. For a sample selected at $i'_{AB}\le 24$, we reach a
  redshift accuracy of $\sigma_{\Delta z/(1+z)}=0.037$ with
  $\eta=3.7\%$ of catastrophic errors (defined strictly as those
  objects with $\Delta z/(1+z) > 0.15$). The reliability of our
  photometric redshifts is lower for fainter objects: we find
  $\sigma_{\Delta z/(1+z)}=0.029, 0.043$ and $\eta=1.7\%, 5.4\%$ for
  samples selected at $i'_{AB}=17.5-22.5$ and $22.5-24$
  respectively. We find that the photometric redshifts of starburst
  galaxies in our sample are less reliable: although these galaxies
  represent only 18\% of the spectroscopic sample they are responsible
  for 54\% of the catastrophic errors. An analysis as a function of
  redshift demonstrates that our photometric redshifts work best in
  the redshift range $0.2\le z \le 1.5$. We find an excellent
  agreement between the photometric and the VIMOS-VLT deep survey
  (VVDS) spectroscopic redshift distributions at $i'_{AB}\le 24$ for
  the CFHTLS-D1 field. Finally, we compare the redshift distributions
  of $i'$ selected galaxies on the four CFHTLS deep fields, showing
  that cosmic variance is already present on fields of $0.7-0.9$
  deg$^2$.  These photometric redshifts will be made publicly
  available from 1st may 2006 at {\it http://terapix.iap.fr} and {\it
  http://cencosw.oamp.fr/}.

   }

\titlerunning{Accurate photometric redshifts for the CFHTLS calibrated using the VVDS}
\authorrunning{Ilbert O. et al.}  
\maketitle

%
%________________________________________________________________

\section{Introduction}

A key factor in the study of galaxy evolution has been our ability to
acquire large, deep, well-defined redshift samples covering
substantial volumes of the Universe. Since the photometric redshift
measurement relies only on the measurement of observed colours
(\cite{Baum62}1962), this technique can be an efficient way to
assemble large and faint samples of galaxies extending to high
redshift. Moreover, the photometric redshift method is also the only
way to estimate redshifts beyond the spectroscopic limit
(\cite{Sawicki97}1997, \cite{Arnouts99}1999, \cite{Benitez00}2000,
\cite{Fontana00}2000, \cite{Bolzonella02}2002).

However, this greatly increased redshift-gathering capability comes at
at price, namely their much lower accuracy with respect to
spectroscopic measurements. The most accurate photometric redshifts
with medium band filters (\cite{Wolf04}2004) still remains around
thirty times less accurate than redshifts measured with low resolution
spectroscopy (\cite{LeFevre04b}2004b). Despite this, for many studies
of the galaxy population, such the galaxy luminosity function, the
velocity accuracy of photometric redshifts is sufficient
(\cite{Wolf03}2003).

To first order, photometric redshifts are reliable when the Balmer or
Lyman continuum breaks can be observed between two broad band
filters. Conventional optical filters from $B$ to the $I$ bands can
therefore measure redshifts between $0.2<z<1$. In addition, near
infrared data are required to provide robust photometric redshifts in
the ``redshift desert'' at $z>1.5$ since the Balmer break is
redshifted to $\lambda>10000\rm{\AA}$ (\cite{Cimatti02}2002,
\cite{Gabasch04}2004, \cite{Mobasher04}2004). Beyond $z>3$, reliable
photometric redshifts can be estimated using deep $U$ or $B$ band
data, based on the Lyman break visible at $\lambda>3600\rm{\AA}$
(e.g. \cite{Madau95}1995).

The reliability of photometric redshifts is also related to the
photometric redshift method. In the standard $\chi^2$ minimisation
method, the most likely redshift and galaxy type are determined by a
template-fitting procedure, which operates by fitting the observed
photometric data with a reference set of spectral templates (e.g.
\cite{Puschell82}1982). Since no spectroscopic information is
required, this standard $\chi^2$ method can provide redshifts beyond
the spectroscopic limit (\cite{Bolzonella02}2002). An alternative
approach is to use a ``training method'' which can extract information
from the spectroscopic sample to estimate the photometric redshifts.
For example, neural network methods (e.g. \cite{Vanzella04}2004) or an
empirical reconstruction of the redshift-colour relation (e.g.
\cite{Connolly95}1995, \cite{Wang98}1998, \cite{Csabai00}2000).
However, if the training set poorly samples the redshift range these
methods can become unreliable. As a hybrid approach combining the
advantages of both methods, the standard $\chi^2$ method
can be optimised using a spectroscopic sample. For instance, the
initial template set can be optimised (\cite{Budavari00}2000,
\cite{Benitez04}2004) or the spectroscopic redshift distribution could
be introduced as a ``prior'' in a Bayesian fitting procedure
(\cite{Benitez00}2000). Essentially, these techniques use the
spectroscopic information to improve photometric redshift quality.

Until now, however, the major limiting factor in the successful
exploitation of photometric redshifts has been our uncertain knowledge
of just how reliable they actually are. Are there systematic trends
between spectroscopic and photometric redshifts? What fraction of
objects have `catastrophic' errors (difference between photometric and
spectroscopic redshifts largely greater than the expected
uncertainty)? How the photometric redshift reliability is correlated
with the galaxy spectral type and the apparent magnitude? Addressing
these issues in a thorough manner requires both a large, highly
uniform photometric sample free from systematic errors and a large,
deep, spectroscopic sample of object selected in the simplest possible
manner. Previous studies at $z>0.3$ have been concerned either very
deep small surveys or larger surveys but with correspondingly
shallower areas. In all cases, the number of available spectroscopic
redshifts has been small (typically less than $\sim 10^3$
objects). The combination of the high-throughput VIMOS wide-field
spectrograph (\cite{LeFevre03}2003) and the MEGACAM survey camera at
CFHT (\cite{Boulade03}2003) makes it possible to amass a large, deep,
highly uniform photometric and spectroscopic samples.

In this paper we present photometric redshifts measured using the
Canada-France Hawaii Telescope Legacy Survey ``Deep Fields''
catalogues (CFHTLS, {\it http://www.cfht.hawaii.edu/Science/CFHTLS})
processed at the TERAPIX data reduction
centre\footnote{terapix.iap.fr} complemented with shallower VIMOS VLT
Deep Survey multi-colour data (\cite{McCracken03}2003,
\cite{LeFevre04a}2004a). We use the current release `T0003' of the
CFHTLS. We focus on the deep field CFHTLS-D1 (or VVDS-0226-04) for
which 11567 faint selected spectra $I_{AB}\le 24.0$ are available from
the VVDS spectroscopic survey (\cite{LeFevre05a}2005a) and are used
here as a training sample. We then compute photometric redshifts for
all the CFHTLS ``Deep Fields'' D1, D2, D3, and D4 to obtain a large
and deep dataset of 522286 objects at $i'_{AB}\le 25$. Photometric and
spectroscopic data are described in Section~2. Results derived with
the standard $\chi^2$ method are presented in
Section~3. We describe in Section~4 how the standard $\chi^2$
 method can be calibrated using spectroscopic data. The
quality of these calibrated photometric redshifts is described in
Section~5, as a function of redshift, apparent magnitude and spectral
type. In Section~6, we investigate how the combination of different
bands affects the accuracy of our photometric redshifts. We finally
present in Section~7 the photometric redshifts with $i'_{AB}\le 25$ in
the 4 CFHTLS deep fields.  More detailed scientific studies such as
the evolution of the angular correlation function or of the galaxy
luminosity function will be deferred to forthcoming articles.

Throughout the paper, we use a flat lambda cosmology ($\Omega_m~=~0.3$,
$\Omega_\Lambda~=~0.7$) and we define
$h~=~H_{\rm0}/100$~km~s$^{-1}$~Mpc$^{-1}$. Magnitudes are given in the
AB system. Photometric and spectroscopic redshifts are denoted by $zp$
and $zs$; $\Delta z$ represents $zp-zs$.

%__________________________________________________________________

\section{Data description}

\subsection{CFHTLS multi-colour data}

\begin{figure}
\centering
\includegraphics[width=9.cm]{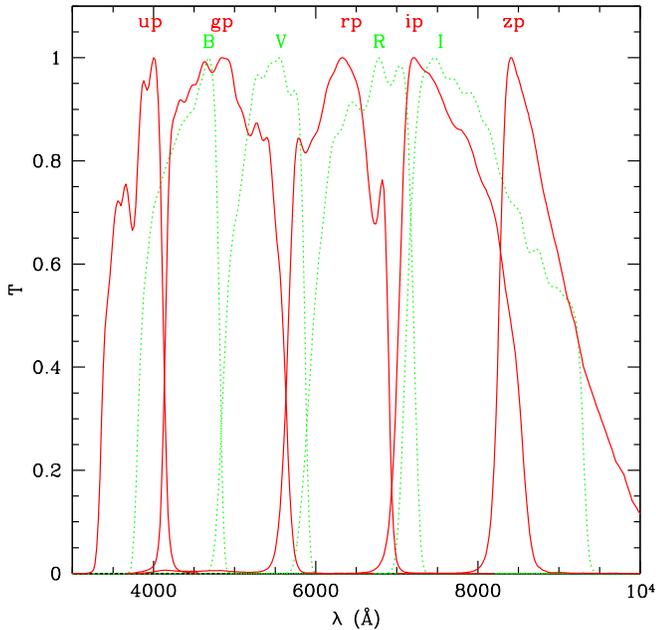}
\caption{CFHT transmissions curves normalised to unity. The solid
  lines correspond to the $u*$, $g'$, $r'$, $i'$, $z'$ MEGACAM filter
  curves; the dotted lines correspond to the $B$, $V$, $R$, $I$
  CFH12K curves.}
\label{filters}
\end{figure}

The MEGACAM deep multi-colour data described in this paper have been
acquired as part of the CFHT Legacy Survey (CFHTLS) which is currently
underway at the 3.6m Canada-France-Hawaii telescope. The MEGACAM
camera consists of 36 CCDs of 2048$\times$4612 pixel and covers a
field-of-view of 1~deg$^2$ with a resolution of 0.186 arcsecond per
pixel. The data covers the observed wavelength range $3500\rm{\AA} <
\lambda< 9400\rm{\AA}$ in the $u^*$, $g'$, $r'$, $i'$, $z'$ filters
(Figure~\ref{filters}). We analyse the four deep CFHTLS fields
CFHTLS-D1 (centred on $02^h25^m59^s-04^\circ 29'40''$), CFHTLS-D2
($10^h00^m28^s+02^\circ 12'30''$), CFHTLS-D3 ($14^h19^m27^s+52^\circ
40'56''$) and CFHTLS-D4 ($22^h15^m31^s-17^\circ 43'56''$), focusing
primarily on the CFHTLS-D1 field for which we have a large
spectroscopic sample available from the VIMOS-VLT deep survey (VVDS).
We use the release `T0003' of the CFHTLS. The data processing of the
CFHTLS ``deep fields'' is described in \cite{McCracken06}(2006, in
preparation). Considerable attention has been devoted in the TERAPIX
pipeline to produce a photometric calibration which is as uniform as
possible over all fields. Comparing the stellar locus in colour-colour
planes in the final four stacks indicates the variation in absolute
photometric zero points field-to-field is less than 0.03 magnitudes
(\cite{McCracken06}2006, in preparation).

After removing the masked area, the effective field-of-view is about
0.79, 0.80, 0.83 and 0.77~deg$^2$ for CFHTLS-D1, D2, D3 and D4
respectively. In CFHTLS-D1, the catalogue reaches limiting magnitudes
of $u^*_{AB} \sim 26.5$, $g'_{AB} \sim 26.4$, $r'_{AB} \sim 25.0$,
$i'_{AB} \sim 25.9$ and $z'_{AB} \sim 25.0$ (corresponding to the
magnitude limit at which we recover 50\% of simulated stellar sources
added to the images using our default detection parameters). The data
in other CFHTLS ``Deep Fields'' are also extremely deep with a
limiting magnitude $i'_{AB} \sim 25.7, 26.2, 26.0$ in the D2, D3, D4
respectively. A summary table listing the exposure times in each band
is given on the TERAPIX web page ({\it http://terapix.iap.fr/}).
Apparent magnitudes are measured using Kron-like elliptical aperture
magnitudes (\cite{Kron80}1980). The magnitudes are corrected from the
galactic extinction estimated object by object from dust map images
(\cite{Schlegel98}1998). We multiply all the SExtractor
(\cite{Bertin96}1996) flux error estimates by a factor of 1.5 to
compensate for the slight noise correlation introduced by image
re-sampling during the stacking of CFHTLS exposures.

\subsection{VVDS multi-colour data}

In addition to CFHTLS data on the CFHTLS-D1 field, we use the shallower
images from the VVDS survey acquired with the wide-field 12K mosaic
camera on the CFHT (\cite{LeFevre04a}2004a). McCracken et al. (2003)
describe in detail the photometry and the astrometry of the
VVDS-0226-04 field. The VVDS-0226-04 field covers the entire CFHTLS-D1
deep field and reaches the limiting magnitudes $B_{AB} \sim 26.5$,
$V_{AB} \sim 26.2$, $R_{AB} \sim 25.9$ and $I_{AB} \sim 25.0$
(corresponding to 50\% completeness). Near infrared data in $J$ and
$Ks$ bands are also available over 160~arcmin$^2$ with the magnitude
limits of $J_{AB}\sim 24.1$ and $K_{AB} \sim 23.8$ respectively
(\cite{Iovino05}2005).\\

\subsection{VVDS spectroscopic data}

We use the VVDS spectroscopic data acquired with the VIsible
Multi-Object Spectrograph (VIMOS) installed at the ESO-VLT. In this
paper, we consider the deep spectroscopic sample observed in the
VVDS-0226-04 field (CFHTLS-D1) and selected according to the criterion
$17.5 \le I_{AB} \le 24.0$ (\cite{LeFevre05a}2005a). This sample
comprises 11567 spectra. Four classes have been established to
represent the quality of each spectroscopic redshift measurement,
corresponding to confidence levels of 55\%, 81\%, 97\% and 99\%
respectively (\cite{LeFevre05a}2005a). Since our goal is to assess the
quality of the photometric redshifts including the fraction of
catastrophic failures, we restrict ourselves to the classes with a
confidence level greater or equal to 97\% (class 3 and 4). In the
sub-area in common with the CFHTLS-D1 field, the final spectroscopic
sample used in this paper consists in 2867 galaxies and 364 stars with
highly reliable redshift measurements. The median redshift is about
0.76. The 1$\sigma$ accuracy of the spectroscopic redshift
measurements is estimated at 0.0009 from repeated VVDS observations.\\

\subsection{Summary}

\begin{figure*}
  \centering \includegraphics[width=15cm]{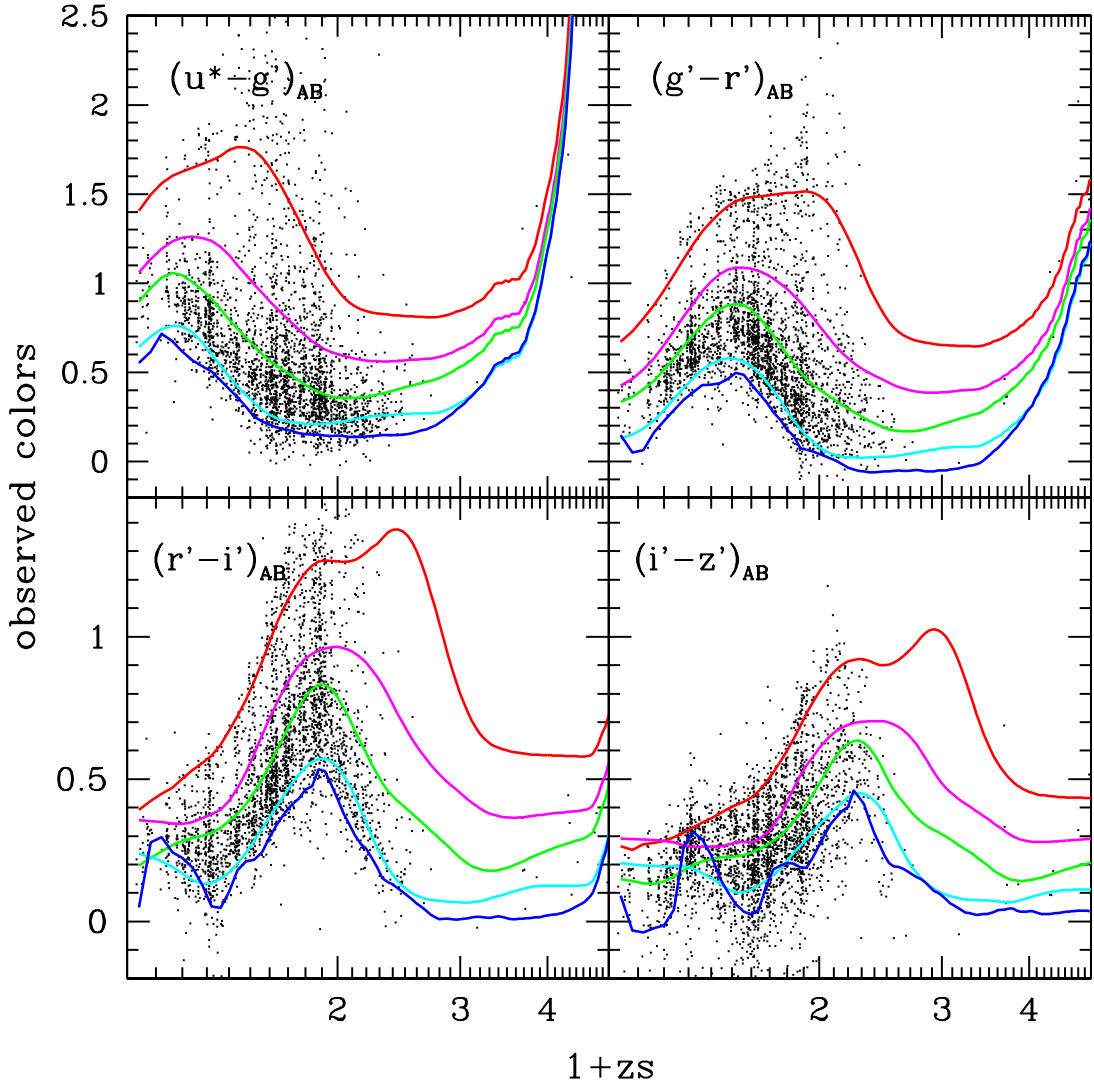}
\caption{Observed colours as a function of the 
  spectroscopic redshifts (black points). The predicted colours
  derived from our optimised set of templates (see section 4.2) are
  shown with solid lines: Ell (red), Sbc (magenta), Scd (green), Irr
  (cyan) (\cite{Coleman80}1980) and starburst (dark blue)
  (\cite{Kinney96}1996) from the top to the bottom, respectively.}
\label{colour_z}
\end{figure*}

To summarise, the multi-colour data on the CFHTLS-D1 field consists in
two joint $u*$, $g'$, $r'$, $i'$, $z'$ and $B$, $V$, $R$, $I$ datasets
over 0.79~deg$^2$, adding also $J$ and $K$ apparent magnitudes over
160~arcmin$^2$. For each object, the photometric redshift is computed
using all the available bands. These photometric redshifts are
calibrated using 2867 spectroscopic redshifts which have a confidence
level greater or equal to 97\%. As an illustration of our combined
photometric and spectroscopic data set, Figure~\ref{colour_z} shows
the observed colours (using only CFHTLS filters) as a function of the
spectroscopic redshifts. Multi-colour data in $u^*g'r'i'z'$ filters is
also available on the other four fields.

\section{Photometric redshifts with the standard $\chi^2$ method}

We present in this Section the results obtained with a standard
$\chi^2$ method, without training the photometric
redshift estimate on the spectroscopic sample.

\subsection{The photometric redshift code {\it Le\_Phare}}

We use the code {\it Le
Phare}\footnote{www.lam.oamp.fr/arnouts/LE\_PHARE.html} (S. Arnouts \&
O. Ilbert) to compute photometric redshifts. The standard $\chi^2$
 method is described in \cite{Arnouts99}(1999,
2002). These photometric redshifts have been found to agree well with
computations from "Hyperz" (\cite{Bolzonella00}2000).

The observed colours are matched with the colours predicted from a set
of spectral energy distribution (SED). Each SED is redshifted in steps
of $\Delta z=0.04$ and convolved with the filter transmission curves
(including instrument efficiency). The opacity of the inter-galactic
medium (\cite{Madau95}1995) is taken into account. The merit function
$\chi^2$ is defined as
\begin{equation}
\chi^2(z, T, A)=\sum_{f=1}^{N_f} \left( \frac{ F_{\rm obs}^f-A\times
F_{\rm pred}^f(z, T)}{\sigma_{\rm obs}^f} \right)^2,
\label{chi2}
\end{equation}
where $F_{{\rm pred}}^f(T, z)$ is the flux predicted for a template
$T$ at redshift $z$. $F_{{\rm obs}}^f$ is the observed flux and
$\sigma_{\rm obs}^f$ the associated error. The index $f$ refers to the
considered filter and $N_f$ is the number of filter. The photometric
redshift is estimated from the minimisation of $\chi^2$ varying the
three free parameters $z$, $T$ and the normalisation factor $A$.

\subsection{Template set}

Our primary template set are the four Coleman Wu and Weedman (CWW)
observed spectra: Ell, Sbc, Scd, Irr (\cite{Coleman80}1980) commonly
used to estimate the photometric redshifts (\cite{Sawicki97}1997,
\cite{Fernandez99}1999, \cite{Arnouts99}1999, \cite{Brodwin06}2006).
We add an observed starburst SED from \cite{Kinney96}(1996) to make
our template sets more representative. These templates are linearly
extrapolated into ultraviolet ($\lambda < 2000 \rm{\AA}$) and
near-infrared wavelengths using the GISSEL synthetic models
(\cite{Bruzual03}2003).  For spectral types later than Sbc, we
introduce a reddening $E(B-V)=0,0.05,0.1,0.15,0.2$ which follows the
interstellar extinction law measured in the Small Magellanic Cloud
(\cite{Prevot84}1984). Even if these five templates are not completely
representative of the variety of observed spectra, it does reduce the
possible degeneracies between predicted colours and redshift
(\cite{Benitez00}2000).

\subsection{Results based on the standard $\chi^2$ method}

\begin{figure}
\centering
\includegraphics[width=9.cm]{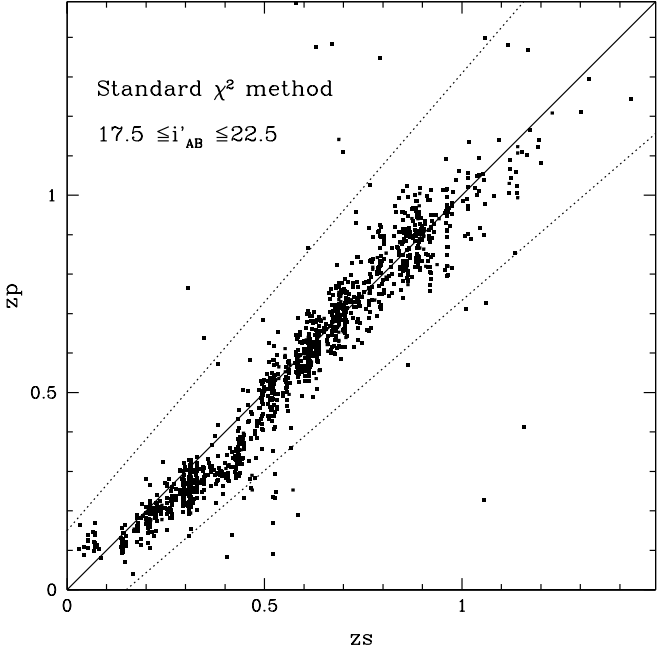}
\caption{Comparison between spectroscopic and photometric redshifts
  determined with the standard $\chi^2$ method (without
  adding the spectroscopic information) for a bright selected sample
  $17.5 \le i'_{AB} \le 22.5$.}
\label{zp_zs_standard}
\end{figure}

We first apply the standard $\chi^2$ method on the
CFHTLS-D1 data without incorporating any spectroscopic information.
Figure~\ref{zp_zs_standard} shows a comparison between the VVDS
spectroscopic redshifts and the photometric redshifts at $i'\le 22.5$.
A clear systematic offset is visible at $zs<0.5$. We would not expect
such a trend to appear for such a relatively bright sample in a
redshift range where the Balmer break is between our $u^*$ and the
$r'$ filters. Small uncertainties in the photometric zero-point
calibration or an imperfect knowledge of the complete instrument
transmission curve are probably responsible for this trend.

\begin{figure}
\centering
\includegraphics[width=9.cm]{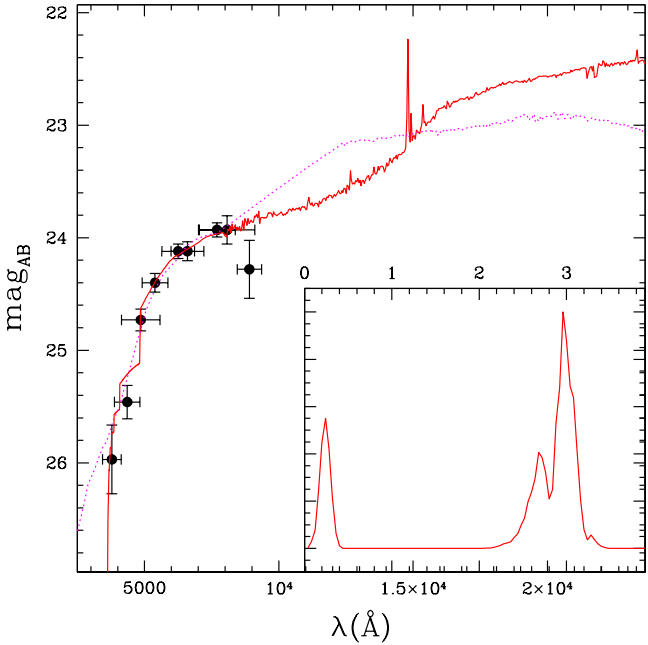}
\caption{Example of best-fitted templates on multi-colour data for a
  galaxy at $zs=0.311$. The solid black points correspond to the
  apparent magnitudes in the $u^*, B, g', V, r', R, i', I, z'$ filters
  from the left to right respectively. The solid line corresponds to a
  template redshifted at $zp=2.97$ and the dotted line at $zp=0.24$.
  The enclosed panel is the associated Probability Distribution
  Function (PDFz).}
\label{2peak}
\end{figure}

At fainter magnitudes (top left panel of Figure~\ref{zp_zs_imp},
method {\it a)}), we see there are a large number of galaxies with
$\Delta z >1$, mainly in the redshift range $1.5<zp<3$. Most of these
catastrophic errors are caused by mis-identification of Lyman and
Balmer break features. An illustration of this degeneracy is presented
in Figure~\ref{2peak}, which demonstrates the importance of
near-infrared data to break this degeneracy. An alternative solution
is to include a relevant information in the redshift probability
distribution function (PDFz) using the Bayesian approach
(e.g. \cite{Benitez00}2000,
\cite{Mobasher04}2004) in order to favour one of the two solutions, as
is discussed in Section \ref{bay}.

This basic comparison shows that blindly trusting the accuracy of
photometric redshifts is perilous. In the following, we will improve
the photometric redshift quality using a spectroscopic training set.

\section{An improved method to compute robust photometric redshifts}

As we demonstrated in Section~3.3, spectroscopic redshifts are
required to calibrate the standard $\chi^2$ method. In
this Section, we describe the steps we have followed to calibrate the
$\chi^2$ photometric redshift estimate.

\subsection{Systematic offsets}

We first select a control sample of 468 very bright galaxies ($i'_{AB}
\le 21.5$) which have spectroscopic redshifts. Using a $\chi^2$
minimisation (equation \ref{chi2}) at fixed redshift, we determine for
each galaxy the corresponding best-fitting CWW template. We note in each
case $F_{obs}^f$ the observed flux in the filter $f$ and $F_{pred}^f$
the predicted flux derived from the best-fit template. For each filter
$f$, we minimise the sum
$$ 
\psi^2=\sum_{i'\le 21.5}^{Ngal} \left( (F_{\rm pred}^f-F_{\rm obs}^f+s^f)/\sigma_{\rm obs}^f \right)^2
$$ leaving $s^f$ as a free parameter. For normal uncertainties in the
flux measurement, the average deviation $s^f$ should be $0$. Instead,
we observe some systematic differences which are listed in
Table~\ref{shift}. Such differences have already been noted by
\cite{Brodwin06}(2006) in the Canada-France Deep Fields Survey. In our
data, these differences never exceed 0.1 magnitude and have an average
amplitude of 0.042 magnitude. They depend weakly on the magnitude cut
adopted to select the bright sub-sample (Table~\ref{shift}) and are
also weakly depending on the set of templates (see Table~\ref{shift}
with the values obtained using the synthetic library PEGASE
\cite{Fioc97}1997). Uncertainties in the calibration of the photometric
zero-points may create discontinuities not reproduced by the templates.
The size of these systematic differences are compatible with the
expected uncertainties in the absolute zero-point calibration (0.05
magnitudes).

We then proceed to correct the predicted apparent magnitudes from
these systematic differences. $s^f$ is the estimated correction that
we apply to the apparent magnitudes in a given filter $f$. If we
repeat a second time the procedure of template-fitting after having
adjusted the zero-points, the best-fit templates may change. We check
that the process is converging: after three iterations the estimated
corrections $s^f$ vary less than 2\%. The values listed in the
Table~\ref{shift} are measured after 3 iterations. Differently from
\cite{Brodwin06}(2006), the corrections used  to correct the apparent
magnitudes are obtained only on a bright sub-sample ($i'\le 21.5$)
after 3 iterations. Since the uncertainties in the zero-point
calibration are not better than 0.01, we add 0.01 in quadrature to the
apparent magnitude errors.

\begin{figure}
\centering \includegraphics[width=9.cm]{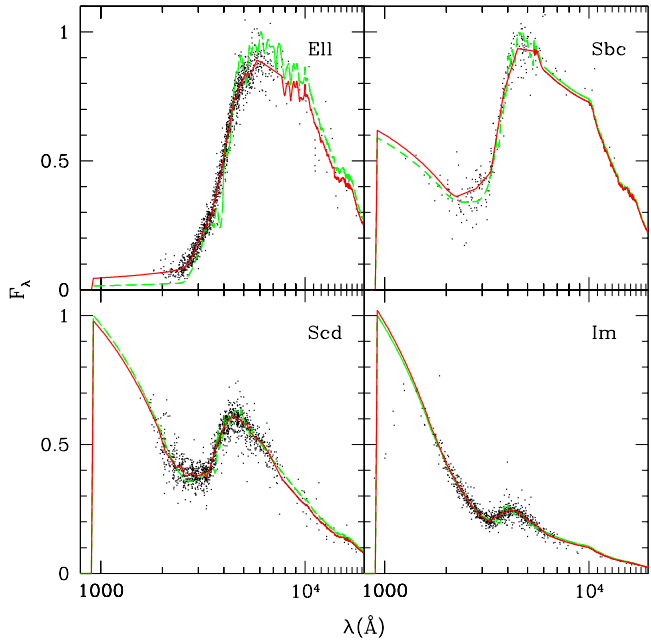}
\caption{Each panel corresponds to one of the four CWW templates (Ell,
  Sbc, Scd, Irr). The points correspond to the flux of each galaxy
  redshifted to the rest-frame using the spectroscopic redshifts. The
  green dashed lines are the initial SEDs and the red solid lines are
  the optimised SEDs which are the output of the procedure described
  in section 4.2.}
\label{templates}
\end{figure}

\begin{table}
\begin{tabular}{c c c c c } \hline \\
        & CWW                & CWW                & CWW                & PEGASE                     \\
filter  & $i'_{AB}<20.5$     &     $i'_{AB}<21.5$ &  $i'_{AB}<22.5$    &   $i'_{AB}<21.5$           \\
\\
\hline\\
$u^*$   &    +0.044 &  +0.045  &  +0.041  &   +0.066           \\
$g'$    &    -0.080 &  -0.080  &  -0.079  &   -0.087           \\ 
$r'$    &    +0.011 &  -0.006  &  -0.012  &   -0.002           \\ 
$i'$    &    -0.005 &  -0.002  &  -0.004  &   -0.001           \\ 
$z'$    &    -0.037 &  -0.025  &  -0.014  &   -0.045           \\ 
$B$     &    +0.057 &  +0.067  &  +0.074  &   +0.063           \\  
$V$     &    -0.066 &  -0.062  &  -0.059  &   -0.066           \\ 
$R$     &    +0.098 &  +0.086  &  +0.083  &   +0.096           \\ 
$I$     &    -0.022 &  -0.001  &  -0.002  &   -0.012           \\

\hline
\end{tabular}
\caption{Systematic differences $s^f$ between observed and predicted
  apparent magnitudes. These values are given for the set of CWW
  templates and for different cuts in apparent magnitudes. We add also
  the values obtained with the synthetic library PEGASE. Throughout
  the paper, we use the values quoted for CWW $i'_{AB}<21.5$.}
\label{shift}
\end{table}

\subsection{Template optimisation}

The apparent magnitude measured in the filter $\lambda_{eff}$ provides
the rest-frame flux at $\lambda_{eff}/(1+z_i)$ for a galaxy with a
spectroscopic redshift $z_i$. Since all the galaxies are at different
redshifts, we can estimate the rest-frame flux over a continuous range
of rest-frame wavelengths from the spectroscopic sample. In this way
can we optimise our set of CWW templates.

We split the galaxy spectroscopic sample according to the best-fit
template (4 CWW + a starburst template with a possible additional
extinction). Keeping only the objects fitted without additional
extinction, we use a sub-sample of 309 galaxies to perform the
template optimisation. The black points in Figure~\ref{templates} show
the rest-frame flux reconstructed from observed apparent
magnitudes. We observe a slight deviation between these points and the
initial templates (dashed lines), particularly for early spectral type
galaxies. We sort the rest-frame flux according to their wavelengths
and bin them by group of 50 points. To produce the optimised
templates, we connect the median flux in each bin (solid lines). When
no data are available, we keep the extrapolation provided by the
initial set of templates. We don't optimise the starburst template to
keep the emission lines in this template.

The colours predicted for these five main optimised templates are
displayed as a function of redshift in Figure~\ref{colour_z}. The
observed trend in the colour-redshift relation are well reproduced by
our set of templates. For $r'-i'$ and $i'-z'$ colours, we observe
oscillations of the predicted colour-redshift relation for the
starburst template. These oscillations are explained by the
contribution of emission lines like H$_{\alpha}$ and OIII to the
observed flux. Since we use only one starburst template to constrain
the possible degeneracies in colour-redshift space, we are not
covering the broad range of possible intensities and line ratios. In
particular, we do not reproduce some blue observed colours
$(r'-i')_{AB}<0.1$ and $(i'-z')_{AB}<0$ (Figure~\ref{colour_z}).  This
lack of representativeness which we have adopted to avoid degeneracies
leads to an accumulation of photometric redshifts in certain peaks of
the colour-redshift relation, for galaxies with strong emission
lines. This is responsible for the presence of narrow peaks in the
redshift distribution for the starburst spectral types.

Finally, these five main optimised templates are linearly interpolated
to produce a total of 62 templates to improve the sampling of the
redshift-colour space and therefore the accuracy of the redshift
measurement.

\subsection{Bayesian approach}
\label{bay}

The Bayesian approach (\cite{Benitez00}2000) allows us to introduce a
relevant {\it a priori} information in the PDFz. Following the
formalism developed by \cite{Benitez00}(2000), we introduce the prior
\begin{equation}
p(z, T | i'_{AB}) = p(T | i'_{AB}) p(z | T,i'_{AB})
\end{equation}                                 
with $p(z | T,i'_{AB})$ the redshift distribution and $p(T | i'_{AB})$
the probability to observe a galaxy with the spectral type $T$.
$p(z|T,i'_{AB})$ is parametrised as:
\begin{equation}                         
p(z|T,i'_{AB}) \propto z^{\alpha_t}exp \left( -\left[
\frac{z}{z_{0t}+k_{mt}(i'_{AB}-20)} \right]^{\alpha_t}\right),
\end{equation}
and $p(T | i'_{AB})$ as:
\begin{equation}
p(T |i'_{AB}) \propto  f_t e^{-k_t(i'_{AB}-20)}.
\end{equation}           
The subscript $t$ denotes the type dependency. Using the formalism
adopted in \cite{Benitez00}(2000), we recompute the values of the free
parameters using the VVDS redshift distribution. We split the sample
according to the four optimised CWW templates. We adjust the parameters
$\alpha_t$, $z_{0t}$, $k_{mt}$ to maximise the likelihood of
``observing'' the VVDS spectroscopic sample. We use the MINUIT package
of the CERN library (\cite{James95}1995) to perform the maximisation
(MIGRAD procedure) and to obtain the corresponding errors (MINOS
procedure).  The values of these parameters for each type are presented
in Table~\ref{paraBen}. The parameters $f_t$ and $k_t$ are also given
in Table~\ref{paraBen} for types one, two, and three and the fraction
of type four is automatically set to complete the sample.

\begin{table*}
\begin{tabular}{c c c c c c} \hline \\
spectral type  &  $\alpha_t$  &  $z_{0t}$   &  $k_{mt}$   &  $f_{t}$   &  $k_{t}$\\
\\
 \hline\\
Ell    & $  3.331^{+  0.109}_{-  0.108}$ & $  0.452^{+  0.015}_{-  0.015}$ & $  0.137^{+  0.007}_{-  0.007}$ & $  0.432^{+  0.047}_{-  0.047}$ & $  0.471^{+  0.043}_{-  0.043}$ \\
Sbc    & $  1.428^{+  0.081}_{-  0.080}$ & $  0.166^{+  0.024}_{-  0.023}$ & $  0.129^{+  0.013}_{-  0.013}$ & $  0.080^{+  0.021}_{-  0.021}$ & $  0.306^{+  0.098}_{-  0.098}$ \\
Scd    & $  1.583^{+  0.038}_{-  0.038}$ & $  0.211^{+  0.015}_{-  0.014}$ & $  0.140^{+  0.006}_{-  0.006}$ & $  0.312^{+  0.033}_{-  0.033}$ & $  0.127^{+  0.036}_{-  0.036}$ \\
Irr    & $  1.345^{+  0.021}_{-  0.021}$ & $  0.204^{+  0.014}_{-  0.014}$ & $  0.138^{+  0.005}_{-  0.005}$ & ... & ... \\
\hline
\end{tabular}
\caption{Parameters used for the prior $P(z,T|i'_{AB})$ using the
  formalism from \cite{Benitez00}(2000). These parameters are derived
  from the VVDS spectroscopic sample.}
\label{paraBen}
\end{table*}

\subsection{Summary}

The photometric redshifts are estimated using the code {\it Le
  Phare}\footnote{www.lam.oamp.fr/arnouts/LE\_PHARE.html} (S. Arnouts
\& O. Ilbert). We calibrate the standard $\chi^2$ method using the VVDS
spectroscopic redshifts:
\begin{itemize}
\item We first adjust iteratively the zero-points of the multi-colour
  catalogue using a bright spectroscopic sample.
\item Then we optimise our primary set of templates using the
  observed flux rest-frame shifted at $\lambda/(1+z_i)$.
\item Finally we apply a prior based on the VVDS redshift distribution
  following the Bayesian formalism presented in \cite{Benitez00}(2000).
\end{itemize}

\section{Results: photometric redshift accuracy}

We now assess the quality of the photometric redshifts obtained with
the calibration method described in Section~4, by comparing the
spectroscopic and photometric redshift samples on the CFHTLS-D1.

\subsection{Method improvement}

\begin{figure*}
\centering \includegraphics[width=17.5cm]{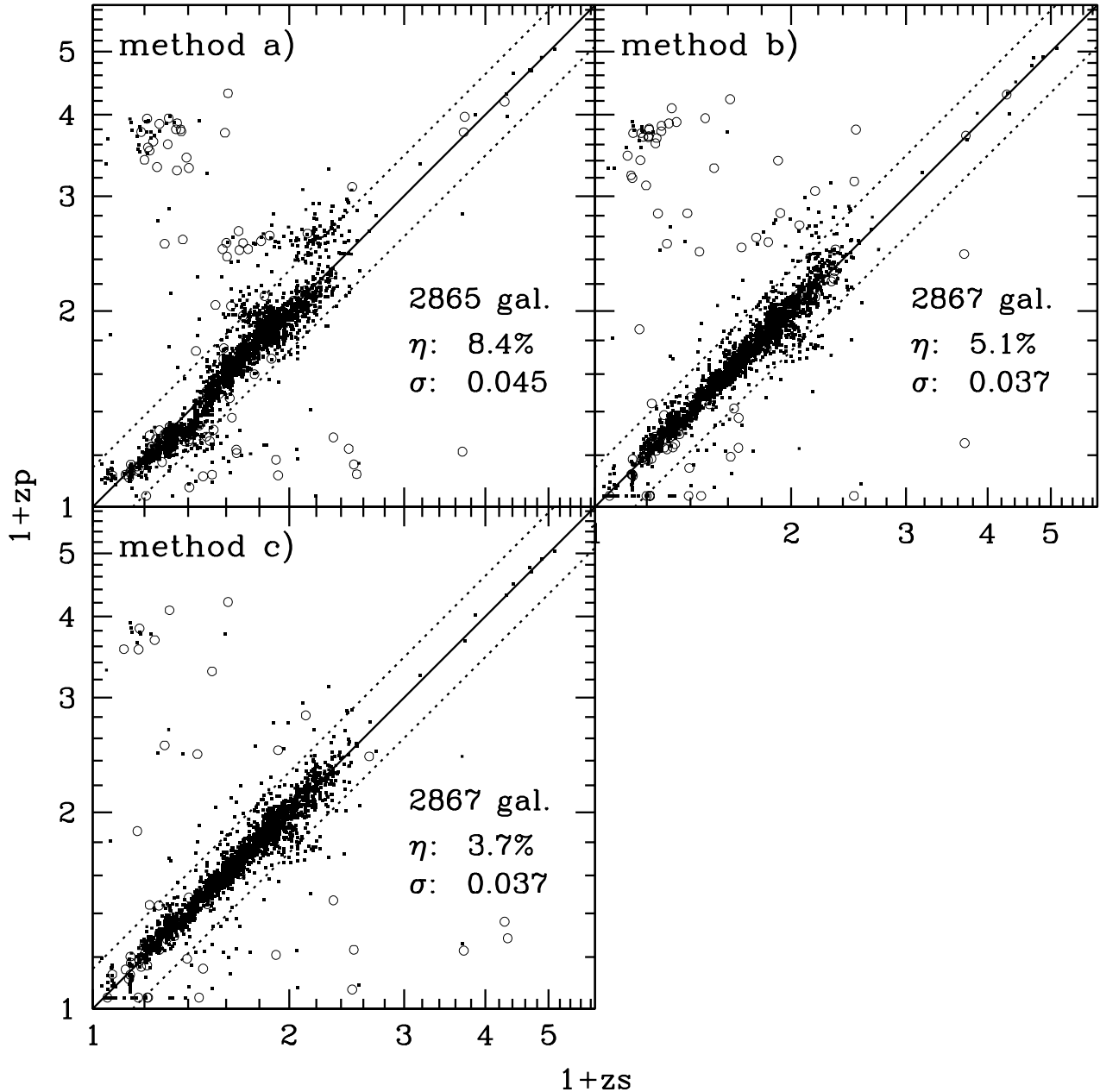}
\caption{Photometric redshifts versus spectroscopic redshifts for the
  sample $17.5<i'_{AB}<24$. Each panel corresponds to an additional
  step in the calibration method with: $method \; a)$ the standard
  $\chi^2$ method ; $method \; b)$ adding the templates
  optimisation and the corrections of the systematic offsets ; $method
  \; c)$ our best method using the Bayesian approach, the templates
  optimisation and the corrections of systematic offsets. The solid
  line corresponds to $zp=zs$. The dotted lines are for $zp=zs \pm
  0.15 (1 + zs)$. We quote as catastrophic errors the fraction $\eta$
  of galaxies with $|zs-zp|/(1+zs)>0.15$ and the accuracy
  $\sigma_{\Delta z/(1+zs)}$. The open symbols correspond to galaxies
  with a second peak detected in the PDFz (probability threshold at
  5\%).}
\label{zp_zs_imp}
\end{figure*}

Figure~\ref{zp_zs_imp} shows the photometric redshifts versus the
spectroscopic redshifts for different steps in the calibration method.
The systematic trends observed with the standard $\chi^2$ method (top
left panel, method {\it a)}) are removed by the template optimisation
and the systematic offset corrections (top right panel, method {\it
b)}). After this step, the accuracy reaches $\sigma_{\Delta
z/(1+zs)}=0.037$. Adding a prior on the redshift distribution
decreases the fraction of catastrophic errors without creating any
systematic trends (bottom left panel, method {\it c)}). The final
fraction of catastrophic errors has decreased by a factor 2.3. In the
following, we restrict our analysis to the best method {\it c)}. This
comparison shows the essential role of the spectroscopic information
to build a robust photometric redshift sample.

With our final calibration method, we reach an accuracy
$\sigma_{\Delta z/(1+zs)}=0.037$. At $i'_{AB}\le 24$, we recover 96\%
of the galaxies in the redshift range $|\Delta z| < 0.15 (1+zs)$.
$\sigma_{\Delta z/(1+zs)}=0.037$ is similar to the accuracy obtained
by the COMBO-17 survey with a larger set of medium band filters
(\cite{Wolf04}2004).  However, considerations on the quality of
photometric redshifts derived from statistical measurements using the
whole sample are not really meaningful since such statistics depends
on the apparent magnitude, the spectral type and the redshift
range. We investigate these dependencies in the next Section.

\subsection{Dependency on apparent magnitude, spectral type and
  redshift}

\begin{figure*}
\centering
\includegraphics[width=17.5cm]{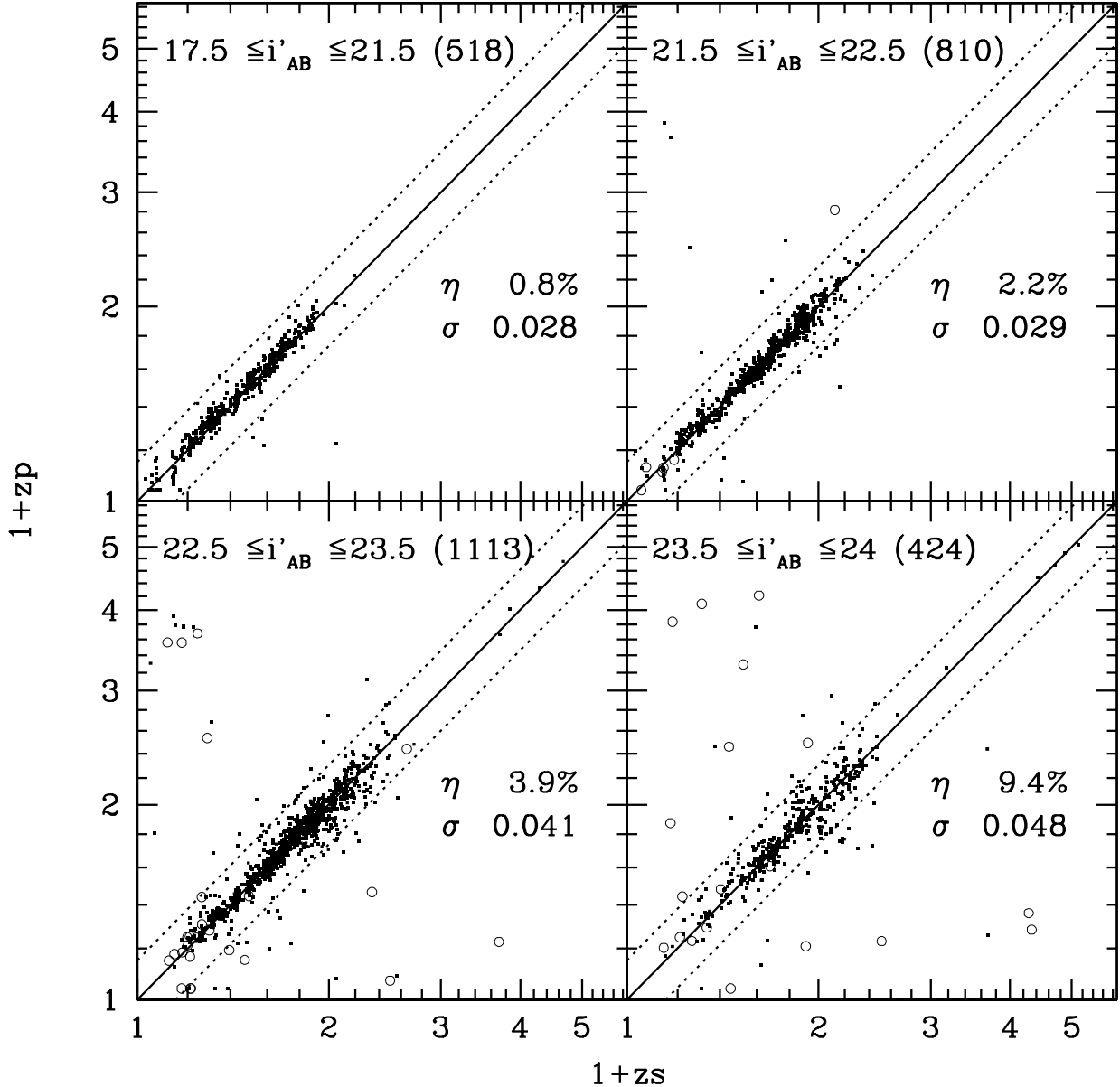}
\caption{Same as Figure~\ref{zp_zs_imp} with the final calibration method
  {\it c)}. Each panel corresponds to a different selection in
  apparent magnitude.}
\label{zp_zs_mag}
\end{figure*}

Figure~\ref{zp_zs_mag} shows the comparison between photometric and
spectroscopic redshifts as a function of apparent magnitude. The
fraction of catastrophic errors $\eta$ increases by a factor of 12
going from $17.5\le i'_{AB} \le 21.5$ up to $23.5 \le i'_{AB} \le 24$.
The redshift rms increases continuously from $\sigma_{\Delta z/(1+zs)}
\sim 0.028$ up to $\sigma_{\Delta z/(1+zs)} \sim 0.048$. The apparent
magnitude is therefore a key parameter, which is to be expected as the
template fitting is less constrained for the fainter objects.

\begin{figure*}
\centering
\includegraphics[width=17.5cm]{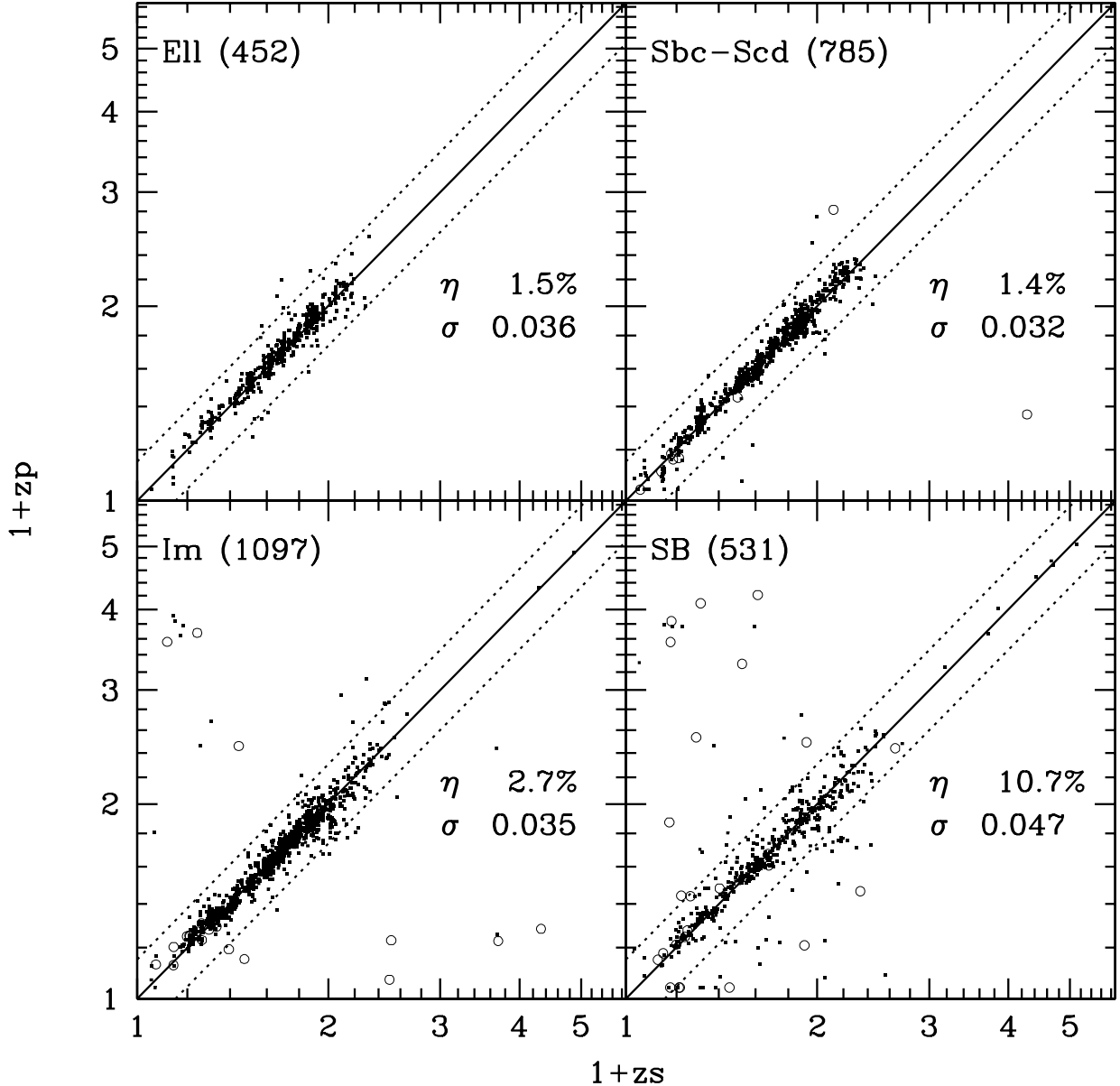}
\caption{Same as Figure~\ref{zp_zs_imp} with our final calibration method
  {\it c)}. The sample is selected at $17.5\le i'_{AB} \le 24$. Each
  panel corresponds to a different selection in spectral type defined
  according to the best-fitting template.}
\label{zp_zs_type}
\end{figure*}

Figure~\ref{zp_zs_type} shows the comparison between photometric and
spectroscopic redshifts as a function of the spectral type. We define
the spectral type according to the best-fit template. The fraction of
catastrophic errors $\eta$ increases by a factor of seven from the
elliptical to the starburst spectral types. The starburst galaxies
represent 18\% of the spectroscopic sample but 54\% of the catastrophic
errors. The accuracy in the redshift measurement is similar for the
Ell, Sbc, Scd and Irr spectral types with $\sigma_{\Delta
  z/(1+zs)}=0.032-0.036$ but rises to $\sigma_{\Delta z/(1+zs)}=0.047$
for the starburst galaxies. Such a dependency on the spectral type is
expected since the robustness of the photometric redshifts relies
strongly on the strength of Balmer break, which is weaker for later
types. In addition, the photometric redshift estimate of late spectral
type galaxies is affected by the intrinsic dispersion in the properties
of the emission lines and by the large range in intrinsic extinction.

\begin{figure}
\centering
\includegraphics[width=9.cm]{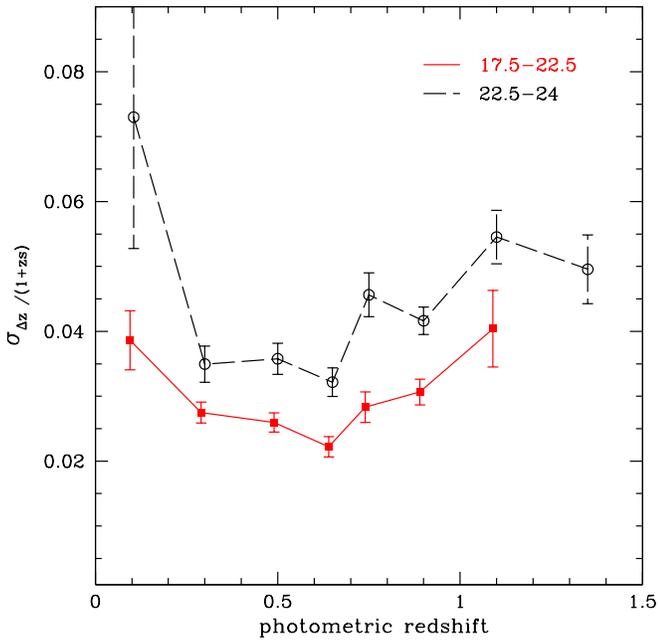}
\caption{Accuracy of the photometric redshifts as a function of
  redshift. Catastrophic errors are removed from the sample. Only bins
  with more than ten objects are shown.}
\label{stat}
\end{figure}

\begin{figure}
\centering
\includegraphics[width=9.cm]{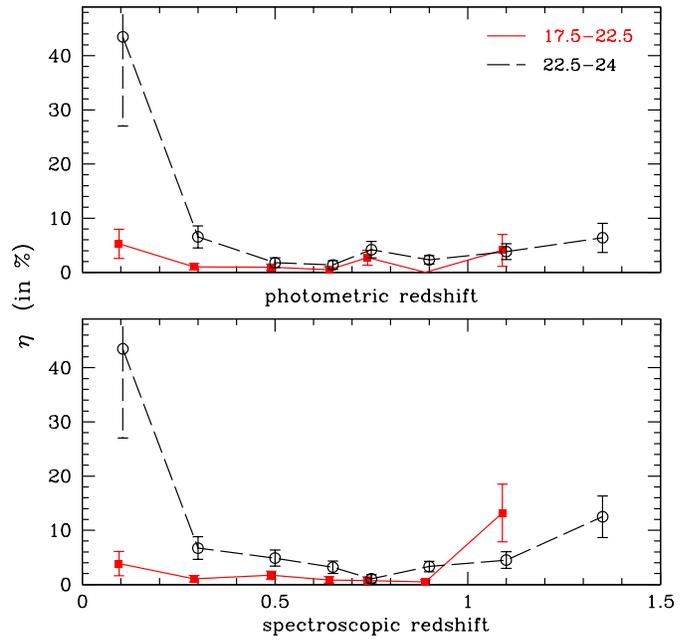}
\caption{Fraction of catastrophic errors $\eta$ per redshift bins. The
  catastrophic errors are defined as galaxies with
  $|zs-zp|/(1+zs)>0.15$. The fraction is measured as a function of the
  photometric redshift (top panel) and of the spectroscopic redshift
  (bottom panel). The top panel shows the level of contamination, i.e.
  the fraction of wrong redshifts in a given photometric redshift
  slice. The bottom panel shows the level of incompleteness, i.e. the
  fraction of redshifts not recovered in a given photometric redshift
  slice. Only bins with more than ten objects are shown.}
\label{cata}
\end{figure}

The photometric redshift reliability also depends on the considered
redshift range. We quantify the dependency on the redshift in
Figure~\ref{stat} and Figure~\ref{cata} showing the rms scatter
$\sigma_{\Delta z/(1+zs)}$ and the fraction of catastrophic errors
$\eta$ as a function of redshift up to $z=1.5$. We split the sample
into a bright $17.5\le i'_{AB}\le 22.5$ and a faint $22.5\le
i'_{AB}\le 24$. We choose the limit $i'_{AB}=22.5$ since it
corresponds to the depth of the shallow VVDS and zCOSMOS spectroscopic
surveys. The fraction of catastrophic errors increases dramatically
only at $z<0.2$.  At $0.2 \le z \le 1.5$, the accuracy is always
better than $0.045(1+zs)/0.55(1+zs)$ for the bright and faint samples
respectively.  The fraction of catastrophic errors $\eta$ remains
always less than $\sim 4\%/14\%$ for the bright and faint sample
respectively (Figure~\ref{cata}). We observe a degeneracy for $zs<0.4$
and $1.5<zp<3$ faint galaxies (bottom left panel of
Figure~\ref{zp_zs_imp}). The origin of this degeneracy is a mismatch
between the Balmer break and the intergalactic Lyman-alpha forest
depression at $\lambda < 1216\rm{\AA}$. 70\% of the galaxies at
$1.5<zp<3$ are in fact at $zs<0.4$ which prevents the use of this
spectral range. At $zs>3$, the Lyman Break is observed between the
$u^*$ band and the $g'$ bands, allowing a reliable photometric
redshift estimate for Lyman Break galaxies. We recover 6 of the 8
galaxies at $zs>3$ (bottom left panel of Figure~\ref{zp_zs_imp}). Even
if the quality of the photometric redshifts appears good, we point out
that we are using only spectroscopic redshifts with the highest
confidence level which is a specific population easier to isolate both
in photometry and in spectroscopy since they have a significant Lyman
break (\cite{LeFevre05b}2005b). Moreover, we have tuned the
calibration method to be efficient at $z<1.5$ using a prior on the
redshift distribution (see Section~4.3) and without allowing galaxies
to be brighter than $M_{B_{AB}}=-24$ (\cite{Ilbert05}2005). We
conclude that the most appropriate redshift range for forthcoming
scientific analysis on the complete population of galaxies is
$0.2<zp<1.5$.

\begin{figure}
\centering \includegraphics[width=9.cm]{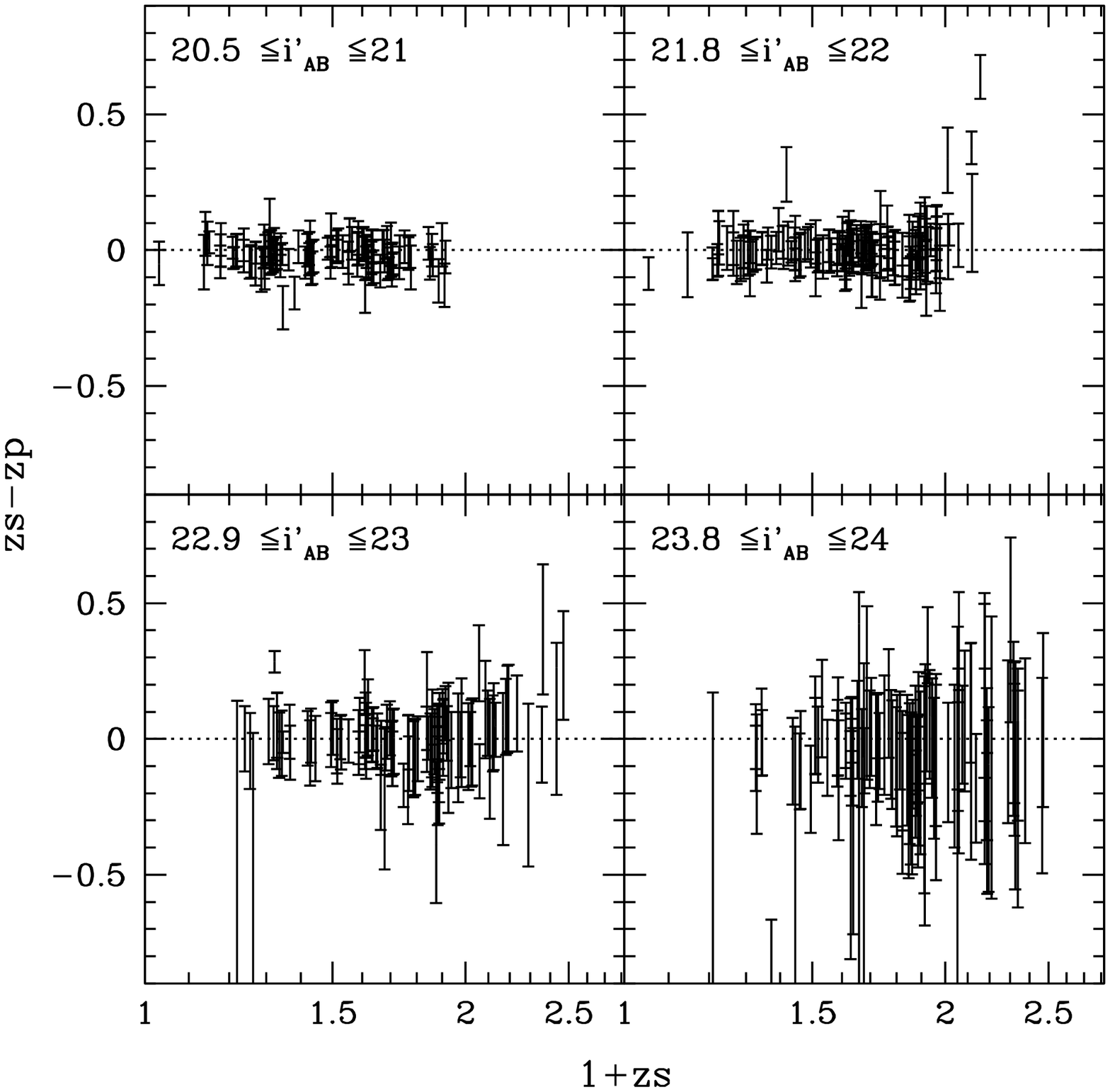}
\caption{$\Delta z$ as a function of redshift in four apparent
  magnitude bins (shown for clarity). We report the
  3$\sigma$ error bars on the photometric redshift estimate.}
\label{errors}
\end{figure}

\subsection{Error analysis}

We investigate here the reliability of the error associated to the
photometric redshift estimate.

The redshift Probability Distribution Function (see
\cite{Arnouts02}2002) is directly derived from the $\chi^2$
distribution
\begin{equation}
PDFz = B \; \exp \left( - \frac{\chi^2(z)}{2} \right),
\end{equation}
with $B$ a normalisation factor. {\it Le\_Phare} (S. Arnouts \& O.
Ilbert) produces the PDFz for each object. A second redshift solution
is likely when a second peak is detected in the PDFz above a given
threshold. An example of galaxy with the good redshift solution
enclosed in the second peak of the PDFz is shown in Figure~\ref{2peak}.
The galaxies with a second peak in the PDFz are flagged with open
circles in Figure~\ref{zp_zs_imp}, Figure~\ref{zp_zs_mag},
Figure~\ref{zp_zs_type} and make up for a large fraction of the
catastrophic errors. We find that the fraction of catastrophic errors
increases dramatically in those cases: when a second peak is detected
with a probability greater than 5\% the fraction of catastrophic errors
increases to $\eta=42\%$. Removing these galaxies from the sample could
be useful to select the most robust sub-sample.

The error bars on the photometric redshifts are given by
$\chi^2(z)=\chi^2_{min}+\Delta \chi^2$. $\Delta \chi^2=1$ and $\Delta
\chi^2=9$ are used to compute the error bars at $1\sigma$ and $3\sigma$
respectively. Figure~\ref{errors} shows the estimated error bars at
3$\sigma$ in narrow bins of apparent magnitudes. The size of the error
bar increases towards faint apparent magnitudes, in a consistent way
with the $\Delta z$ rms. We find that 67\% and 90\% of the
spectroscopic redshifts are well located in the 1$\sigma$ and 3$\sigma$
error bars respectively. We note that these values remain lower than
the theoretical values since photometric uncertainties (such as
blending or the presence of bright neighbours) or the suitability of
our template set are not taken into account in the computation of the
PDFz. We conclude that our 1$\sigma$ error bars are an accurate
representation of the photometric redshift error, and there can be a
useful way to assess their reliability beyond the spectroscopic limit.

\subsection{Comparison between photometric and spectroscopic redshift
  distributions}

In order to calculate the galaxy redshift distribution, we first need
to remove the stars from the sample. We use the half-light radius
$r_{1/2}$, a morphological criterion measured provided by SExtractor
(\cite{Bertin96}1996). From the spectroscopic sample, we find that
95\% of the stars have $r_{1/2}<2.7$. Since 16\% of the galaxies have
also $r_{1/2}<2.7$, we combine this morphological criterion with a
colour criterion. For each object, we compute simultaneously the
$\chi^2$ for the galaxy library and the $\chi^2_s$ for the star
library (\cite{Pickles98}1998). If the conditions $\chi^2-\chi^2_s>0$
and $r_{1/2}<2.7$ are satisfied simultaneously, the object is flagged
as a star. Applying these criteria on the spectroscopic sample, we
recover 79\% of the stars and only 0.77\% galaxies are misclassified
as stars.  The remaining 21\% of stars are misclassified as galaxies
and 69\% of these are in the redshift range $zp<0.2$.

\begin{figure*}
  \centering \includegraphics[width=17cm]{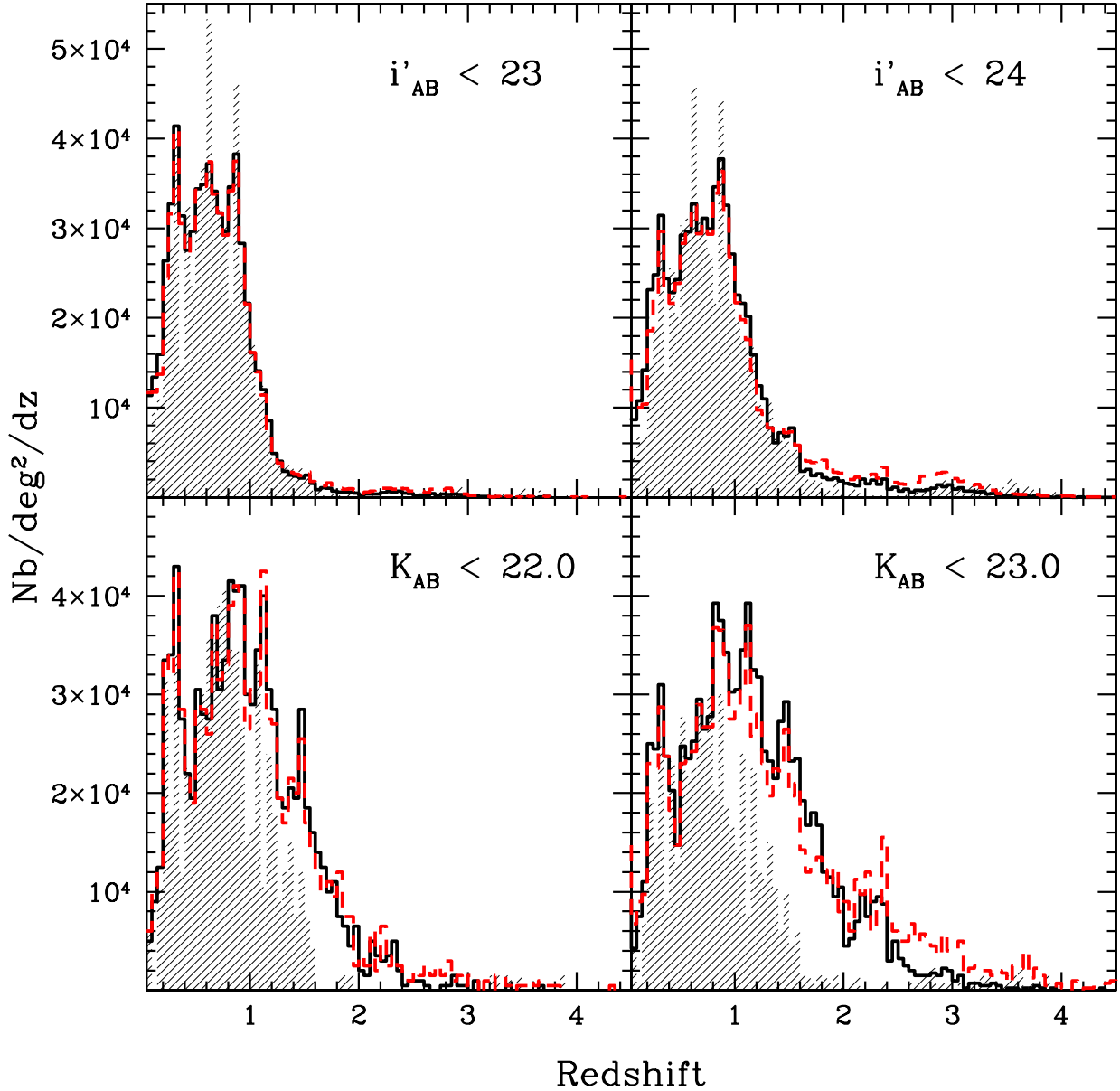}
  \caption{Comparison between the photometric redshift distributions
  and the VVDS spectroscopic redshift distributions on the CFHTLS-D1
  field, for samples selected at $i'_{AB} <23$ (top left), $i'_{AB}
  <24$ (top right), $K_{AB} <22$ (bottom left) and $K_{AB} <23$
  (bottom right). The black solid lines and the red dashed lines
  correspond respectively to the estimate with and without using the
  prior on the redshift distribution. These distributions are compared
  with the spectroscopic redshift distributions (shaded histograms)
  from the VVDS sample, originally selected at $I_{AB}\le 24$. To
  maintain the same vertical axis, the redshifts distributions are
  divided by a factor of two for $i'\le 24$, $K\le 23$.}
\label{z_dist}
\end{figure*}

Since we use the spectroscopic redshift distribution as a prior (see
Section~4.3) a critical point is to check at which level the
photometric redshift distribution depends on the prior. We compare the
redshift distributions obtained using the prior (weighted solid lines)
and without (dashed lines) in Figure~\ref{z_dist}. The prior has no
impact on the global shape of the redshift distribution. We see
significant differences in the redshift distributions at $z>1.5$, where
the prior efficiently removes the catastrophic failures at $1.5<zp<3$.

For the $i'_{AB}\le 23$ and the $i'_{AB}\le 24$ selected samples, we
compare in Figure~\ref{z_dist} the photometric and the VVDS
spectroscopic redshift distributions. The distributions are in
excellent agreement to $z\sim 1.5$. At $1.5<z<3$, the photometric
redshifts are contaminated by low redshift galaxies (see Section~5.2)
and the paucity of spectral features in UV makes spectroscopic
redshift measurement difficult (\cite{LeFevre05a}2005a). Both effects
explain the difference between the photometric and spectroscopic
redshift distribution at $z>1.5$.

We are able to identify peaks in the photometric redshift distribution
which are clearly associated with peaks in spectroscopic redshift
distribution. We smooth the spectroscopic and photometric redshift
distributions using a sliding window with a step $\Delta z=0.2$. The
ratio between the observed redshift distribution obtained with a step
$\Delta z=0.01$ and the smoothed redshift distribution shows three
peaks at $zp \sim 0.31, 0.61, 0.88$ in the photometric redshift
distribution, in excellent agreement with peaks identified at $zs \sim
0.33, 0.60, 0.89$ in the spectroscopic redshift distribution. The
significance of the peaks is lower by a factor two in the photometric
redshift sample since the peaks are broadened by the uncertainties on
the photometric redshift estimates.

\begin{table*}
\begin{tabular}{c c c c c } \hline \\
Magnitude cut & $i^*\le 23$ & $i^*\le 24$ & \hspace{0.5cm} $Ks \le 22$
& $Ks \le 23$ \\ \hline \\ $z_m$ & 0.76 & 0.90 & \hspace{0.5cm} 0.90 &
1.07 \\
\% at $z>1$       &  13\% &    28\%    & \hspace{0.5cm}  28\%  &  43\%   \\
\hline
\end{tabular}
\caption{Median redshifts and fraction of galaxies at $z>1$ for
samples selected according to $i'_{AB}\le
23, 24$ and $K_{AB}\le 22, 23$ in the CFHTLS-D1}
\label{zm}
\end{table*}

\section{Added value of each multi-colour data set}

We now compare the reliability of photometric redshifts computed from
either a $BVRI$ dataset using the VVDS-CFH12K photometry or
$u^*g'r'i'z'$ using the CFHTLS-MEGACAM, and quantify the useful range
of photometric redshifts for each of these datasets. In addition, we
explore the added value of different bands to the accuracy and
reliability of photometric redshifts.

\subsection{Added value of $u^*$ and $z'$ bands}

Removing successively the $u^*$ and the $z'$ bands, we investigate
possible systematic trends if these bands are not available or if they
are shallower.

Figure~\ref{zp_zs_data} (top left-hand panel) shows the photometric
redshifts computed without $u^*$ band data. Only $\sim70\%$ of the
photometric redshifts at $zs<0.4$ are recovered which should be
compared with $\sim90\%$ using the $u^*$ band. Since the filter system
is no longer sensitive to the Lyman break, a large fraction of low
redshift galaxies contaminates the $zp>3$ redshift range. This test
shows the importance of a deep $u^*$ band to constrain the photometric
redshifts at $z<0.4$ and $z>3$.

Figure~\ref{zp_zs_data} (top right panel) shows the photometric redshifts
computed without $z^*$ band data. Most of the photometric redshifts at
$zs>1$ are estimated at $zp \le 1$. We observe an accumulation of
photometric redshifts around $zp \sim 0.8-0.9$. This trend is expected
since the filter system is no longer sensitive to the Balmer break at
$z>1$. In this case the use of the photometric redshifts is problematic
even at $z < 1$ without $z'$ band data.

\subsection{Photometric redshifts from the VVDS imaging survey alone}

The VVDS multi-colour survey was carried out in $B$, $V$, $R$ and $I$
bands over 10 deg$^2$ (\cite{LeFevre04a}2004a) using the CFH12K
wide-field mosaic camera at CFHT. The accuracy of the photometric
redshifts using only BVRI is presented in Figure~\ref{zp_zs_data}
(bottom right panel). Since the VVDS photometric survey is shallower,
the quality of the photometric redshifts is obviously worse than the
results presented previously. $\eta$ rises to 19.1\% which is a factor
of five greater than our best value. As we demonstrate in Section~6.3,
the absence of deep $u^*$ and $z'$ band data makes it difficult to
compute photometric redshifts at $zs<0.4$ and $zs>1$.  However, even
using only four broad bands, we recover 80\% of the spectroscopic
redshifts at $I_{AB}\le 24$ with $\sigma_{|\Delta z|/(1+zs)}=0.057$.

\subsection{Photometric redshifts from the CFHTLS imaging data alone}

\begin{figure*}
\centering
\includegraphics[width=17cm]{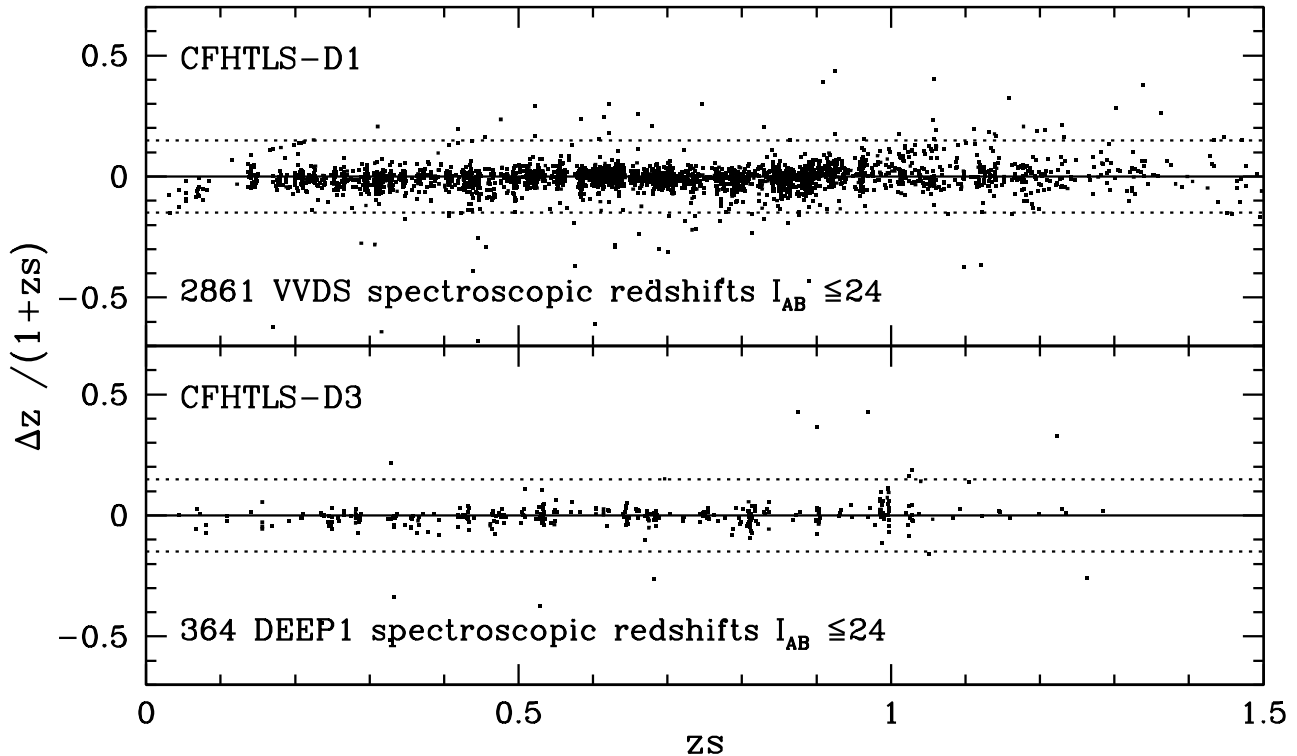}
\caption{$\Delta z$ as a function of redshift. The photometric
  redshifts are computed using the CFHTLS filter set $u^*$, $g'$,
  $r'$, $i'$, $z'$. The top and bottom panels present the
  photometric redshifts obtained on the CFHTLS-D1 and CFHTLS-D3 fields
  respectively.}
\label{zp_zs_CFHTLS}
\end{figure*}

The deep CFHTLS survey consists in four fields imaged over 3.2 deg$^2$
in the $u^*g'r'i'z'$ filters. The quality of the photometric redshifts
computed using only the $u^*g'r'i'z'$ bands is displayed in
Figure~\ref{zp_zs_CFHTLS} (top panel) for the CFHTLS-D1. We find
$\eta=4.2\%$ and $\sigma_{|\Delta z|/(1+zs)}=0.040$. These CFHTLS
photometric redshifts are close to be as accurate as the photometric
redshifts computed using the full photometric dataset. Photometric
redshifts for the other CFHTLS deep fields will be introduced in
Section~7.

\subsection{The near-infrared sample}

\begin{figure*}
  \centering \includegraphics[width=17cm]{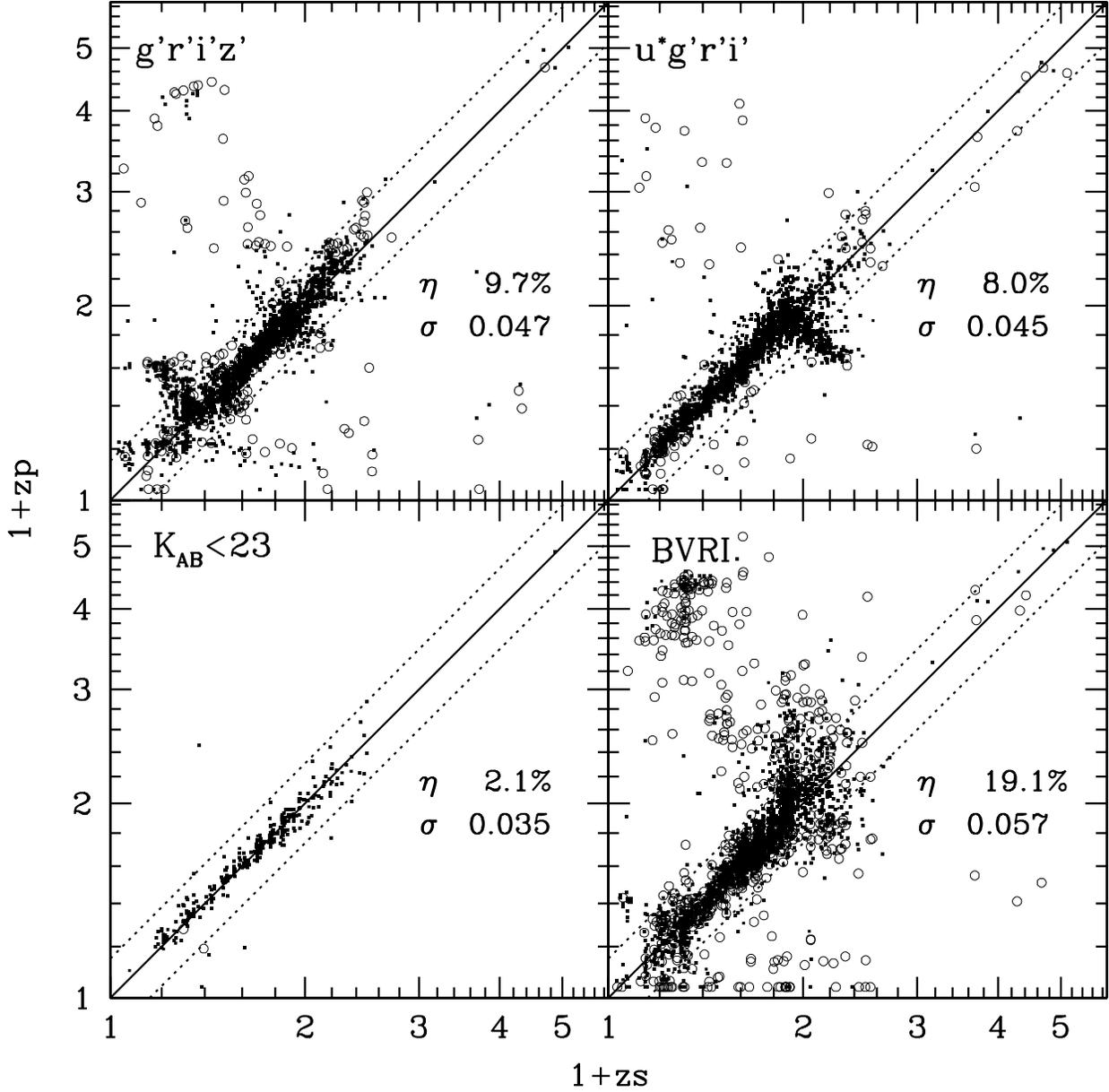}
  \caption{Comparison between spectroscopic and photometric redshifts
    for different combinations of filters. The photometric redshifts in
    the top left and right panel are computed without using the deep
    $u^*$/$z'$ band respectively. The bottom left panel shows the
    photometric redshifts for a near-infrared selected sample computed
    using $B$, $V$, $R$, $I$, $u^*$, $g'$, $r'$, $i$, $z'$, $J$ and $K$
    bands. The bottom right panel shows the photometric redshifts
    obtained using the $B$, $V$, $R$, $I$ bands from the VVDS survey.}
\label{zp_zs_data}
\end{figure*}

Deep near-infrared observations in the $J$ and $K$ bands are available
for a 160~arcmin$^2$ (\cite{Iovino05}2005) sub-area of the D1 field.
This complete sub-sample of 3688 galaxies at $K_{AB}\le 23$ represents a
unique dataset in term of depth and area (it is one magnitude deeper
and covers a three times larger area than the K20 survey,
\cite{Cimatti02}2002). Near-infrared bands are crucial to constrain the
photometric redshifts in the `redshift desert' since the $J$ band is
sensitive to the Balmer break up to $z\sim2.5$ and enters in the $K$
band at $z> 3.8$. The photometric redshifts for galaxies selected at
$K_{AB}<23$ are shown in Figure~\ref{zp_zs_data} (bottom left panel). We
obtain the most reliable photometric redshifts on this sub-sample with
$\eta=2.1\%$ and $\sigma_{|\Delta z|/(1+zs)}=0.035$.

We present in Figure~\ref{z_dist} the redshift distributions for
samples selected with $K_{AB}\le 22$ and $23$. Comparing the
photometric and spectroscopic redshift distributions, we see a large
difference in the high redshift tail which is explained by the colour
incompleteness caused by the $I_{AB}\le 24$ selection function of the
spectroscopic sample. The median redshifts and the fraction of
galaxies with $z>1$ are given in Table \ref{zm}. As expected, we find
that near-infrared selected samples are more efficient to target a high
redshift population than $i'$ selected sample. We find 43\% of the
galaxies at $z>1$ for a sample selected at $K_{AB}\le 23$.  As previous
$K$ selected surveys (\cite{Cimatti02}2002, \cite{Somerville04}2004)
have found, a large population of galaxies at $z>1$ is observed.

\begin{table}
\begin{tabular}{c c c c c } \hline \\
                    &  D1     &       D2     &  D3      &  D4  \\
 \hline \\
$17.5<i'_{AB}<22$   &  0.49   &      0.48    &  0.49    & 0.48     \\        
$22<i'_{AB}<23$     &  0.75   &      0.75    &  0.75    &  0.81    \\        
$23<i'_{AB}<24$     &  0.89   &      0.86    &  0.86    &  0.95    \\        
$24<i'_{AB}<25$     &  1.00   &      0.95    &  0.94    &  1.03   \\        
\hline
\end{tabular}
\caption{Median redshifts in the four CFHTLS deep fields (columns) for samples 
selected according to $17.5<i'_{AB}<22$, $22<i'_{AB}<23$,
$23<i'_{AB}<24$, $24<i'_{AB}<25$ from the top to the bottom,
respectively.}
\label{zm_CFHTLS}
\end{table}

\begin{figure*}
  \centering \includegraphics[width=15cm]{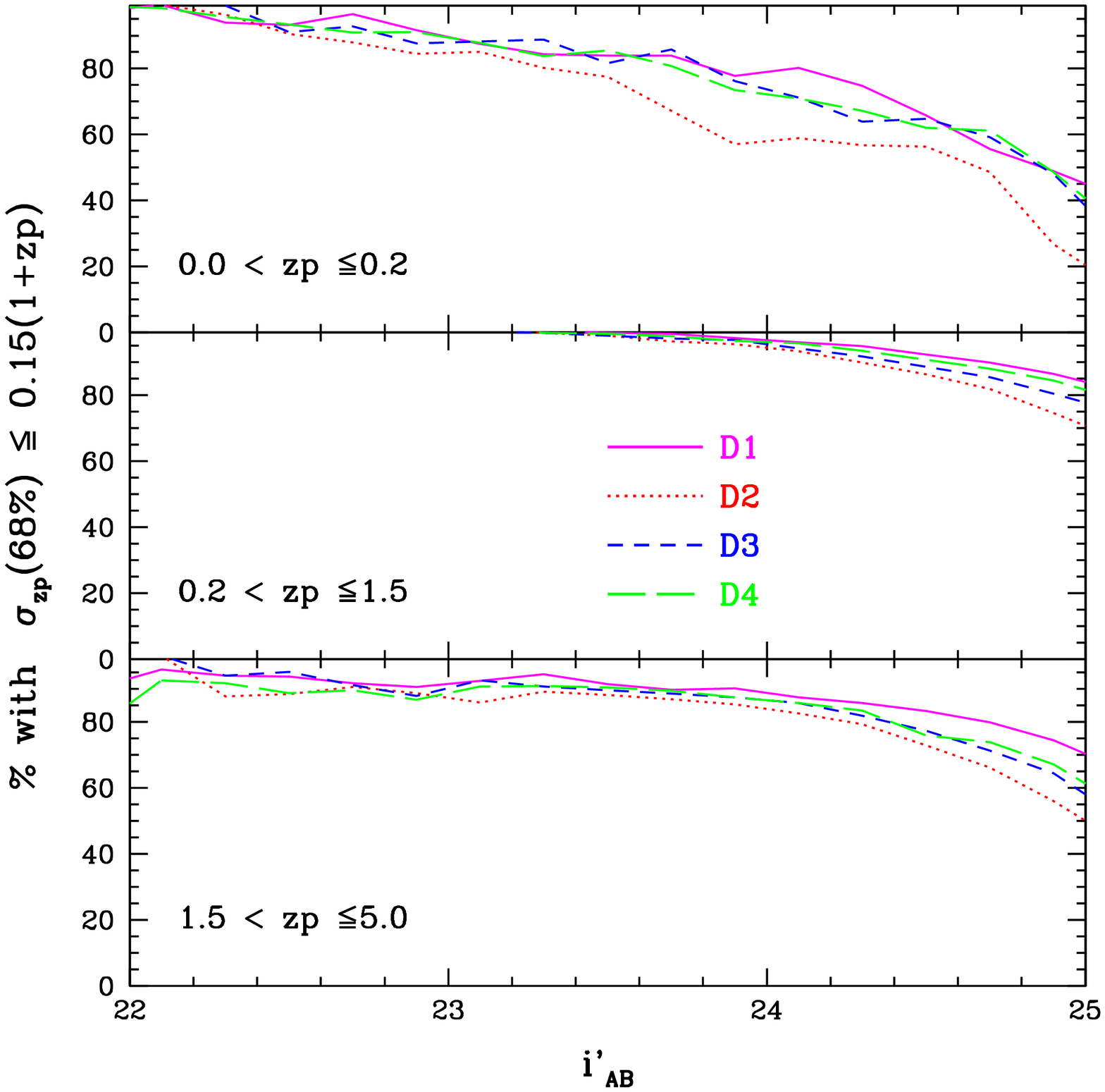}
  \caption{Fraction of photometric redshifts with an error bar 68\%
  smaller than $0.15\times (1+zp)$ as a function of $i'$ apparent
  magnitude. This statistic is shown in the four CFHTLS deep fields:
  D1 (solid line), D2 (dotted line), D3 (short dashed line) and D3
  (long dashed line) and in the three redshift ranges $0<z \le 0.2$
  (top panel), $0.2<z \le 1.5$ (middle panel) and $1.5<z \le 5$
  (bottom panel).}
\label{error_I}
\end{figure*}

\section{Photometric redshifts in the CFHTLS ``Deep Fields'' D1, D2, D3
and D4}

We finally use the photometric redshift calibration derived from the
CFHTLS-D1 field and the VVDS spectroscopic sample to derive
photometric redshifts for all fields of the CFHTLS deep survey.

The three CFHTLS deep fields D2, D3, D4 have been imaged with the same
instrument and are reduced homogeneously in exactly the same way as the
D1 field (\cite{McCracken06}2006, in preparation). We therefore assume
that we can measure the photometric redshifts for these fields in the
same manner as we have for the D1 field. As a consistency check, we use
364 spectroscopic redshifts from the DEEP1 survey
(\cite{Phillips97}1997) which are in the D3 field. This allows us to
test blindly the quality of these photometric redshifts without any
additional calibration. The comparison is shown in the bottom panel of
Figure~\ref{zp_zs_CFHTLS}. We find $\eta=3.8\%$ and $\sigma_{|\Delta
  z|/(1+zs)}=0.035$ at $i'_{AB}\le 24$ and $z<1.5$, without any
systematic trend. We therefore conclude that our calibration method
derived from D1 can be applied to the other CFHTLS deep fields.

The photometric redshift quality using only the $u^*$, $g'$, $r'$,
$i'$, $z'$ has already been discussed in Section~6.3, but only for the
CFHTLS-D1 field and for $i'_{AB}\le 24$. Since we have already
demonstrated in Section~5.3 that the 1$\sigma$ error bars are
representative of a measurement at 68\% of confidence level, we use the
1$\sigma$ error bars to quantify the accuracy of the photometric
redshifts in the different fields and beyond the spectroscopic limit.
Figure~\ref{error_I} shows the fraction of photometric redshifts with a
$1 \sigma$ error bar left than $0.15\times (1+z)$. The best constraint
is obtained on the CFHTLS-D1 field and gradually declines for the D4,
D3 and D2. The constraint on the photometric redshifts is the lowest on
the D2, which is expected since the total exposure times in the $u^*$
and $z'$ bands are respectively 7.7 and 1.7 times lower for the D2
field than for the the D1 field. We note that the specific trends
described in Section~6.1 and shown in Figure~\ref{zp_zs_data} could
partially affect the photometric redshift estimates for the CFHTLS-D2
given than this field has substantially shallower $u^*$ and $z'$ data.
The other significant trends observed in Figure~\ref{error_I} are
expected from our previous comparisons:
\begin{itemize}
\item the redshift range $0.2 < z \le 1.5$ is the most suitable
  for the 4 fields which is expected since this redshift range is
  constrained by the set of filters used.
\item the accuracy of the photometric redshifts decreases toward
  fainter apparent magnitudes, faster at $i'_{AB}>24$. For
  $0.2<z<1.5$, the fraction of galaxies with
  $\sigma_{zp}(68\%)>0.15\times (1+z)$ remains greater than $\sim 80\%$
  at $i'_{AB}=25$ in the CFHTLS-D1 field.
\end{itemize}

\begin{figure*}
\centering
\includegraphics[width=15cm]{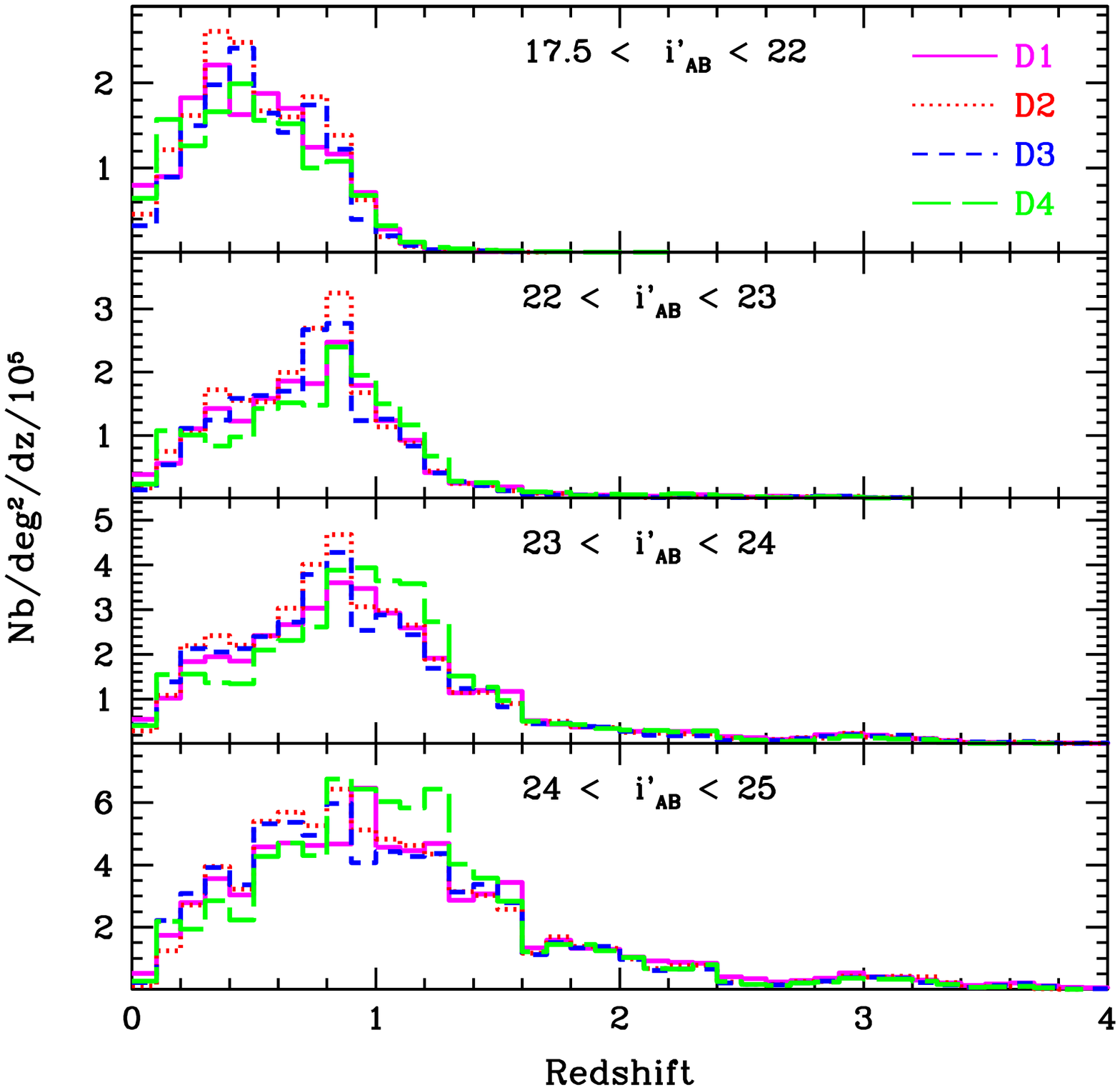}
\caption{Photometric redshift distributions in the 4
  fields CFHTLS-D1 (solid line), CFHTLS-D2 (dotted line), CFHTLS-D3
  (short dashed line) and CFHTLS-D4 (long dashed line). The redshift
  distribution are shown from bright ($17.5 < i'_{AB} <22$) to faint
  selected samples ($24 < i'_{AB} <25$) from the top to the bottom,
  respectively.}
\label{distz_CFHTLS}
\end{figure*}

\begin{figure}
\centering
\includegraphics[width=9cm]{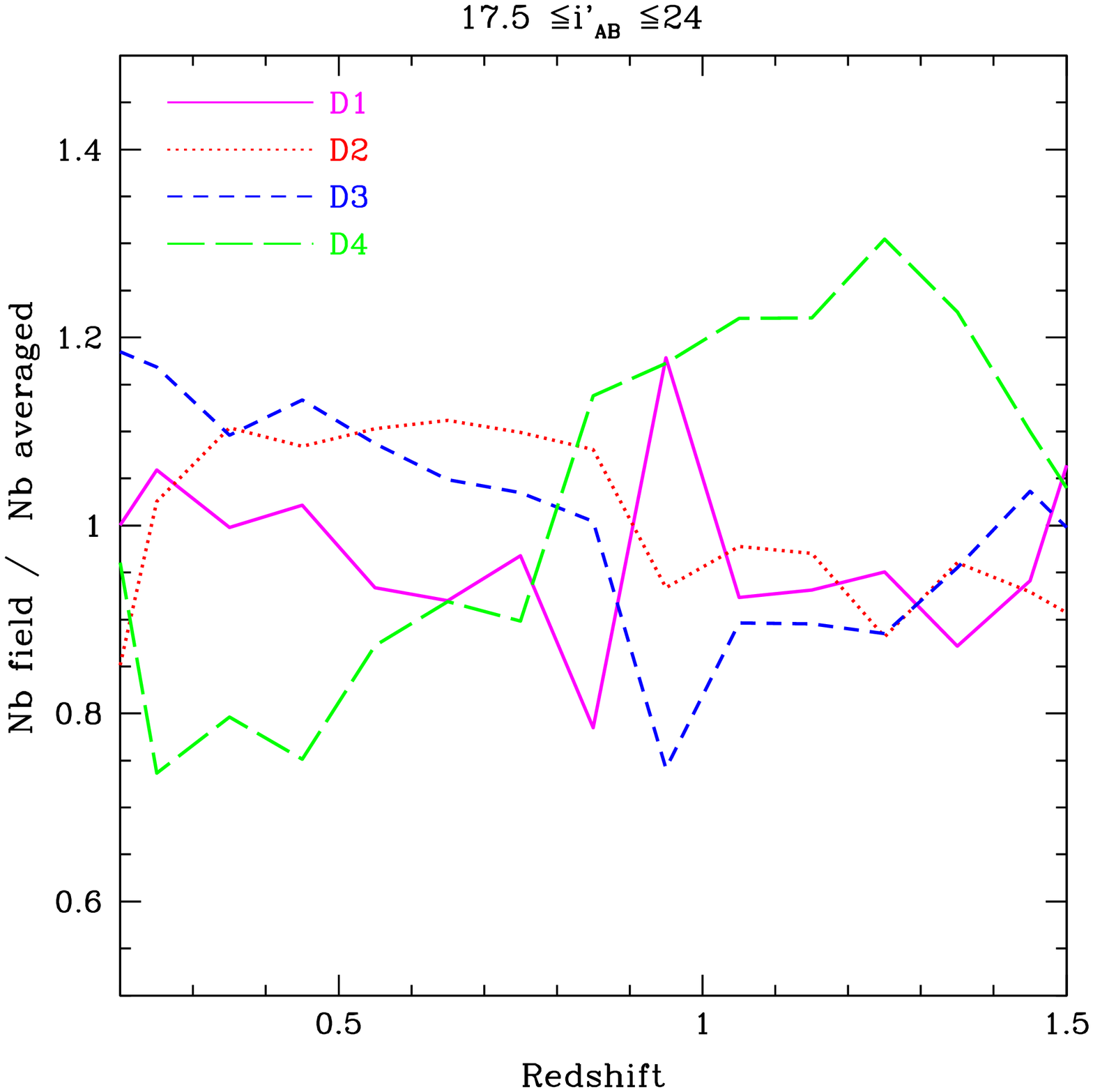}
\caption{Ratio between the redshift distributions in each field and the 
redshift distribution averaged over the 4 fields. Symbols for each
field are the same than Figure~\ref{distz_CFHTLS}.}
\label{CV}
\end{figure}

We show in Figure~\ref{distz_CFHTLS} the redshift distributions for
the four CFHTLS deep fields. As expected, the median redshift
increases for fainter samples (Table~\ref{zm_CFHTLS}) rising from
$z_{m}\sim 0.45$ at $i'_{AB}\le 22$ to $z\sim 1$ at $24\le i'_{AB} \le
25$. The median redshifts are in good agreement between the four
fields. The redshift distribution in the D4 is shifted at higher
redshift. We observe significant variations of the redshift
distribution between the four fields. Figure~\ref{CV} shows the ratio
between the redshift distribution in each field and the redshift
distribution averaged over the four fields, using a redshift step of
$\Delta z=0.1$. This ratio shows that the difference in redshift
distribution can reach at most a factor 1.6 in a redshift bin
$\Delta=0.1$ (at z=0.25 between the D2 and the D4 fields). The average
dispersion in the interval $0.2<z<1.5$ and in a redshift slice
$\Delta=0.1$ is $\sim 15\%$. We conclude that the cosmic variance can
be important for fields of $\sim$0.8 deg$^2$. We note however that the
four fields do not reach exactly same depth in all filters which could
be responsible for some of the differences between the redshifts
distributions.

\section{Conclusions}

Using the unique combination of the deep $u^*g'r'i'z'$ multi-band
imaging data from the CFHTLS survey supplemented by shallower $BVRI$
data from the VVDS imaging survey (and also by $J$ and $K$ data on a
smaller sub-area) and VVDS first epoch spectroscopic redshifts, we have
been able to obtain very accurate photometric redshifts on the
CFHTLS-D1 field. We reach $\sigma_{\Delta z/(1+z)}=0.037$ at $i'\le 24$
and $\eta=3.7\%$ of catastrophic errors (defined strictly as $\Delta z
> 0.15(1+z)$). For the bright sample selected at $i_{AB}\le 22.5$, we
reach $\sigma_{\Delta z/(1+z)}=0.030$ and $\eta=1.7\%$.

This accuracy has been achieved by calibrating our photometric
redshifts using a large and deep spectroscopic sample of 2867 galaxies.
We have established a reliable calibration method combining an
iterative correction of photometric zero-points, template optimisation,
and a Bayesian approach. This method removes some obvious systematic
trends in the estimate and reduces by a factor 2.3 the fraction of
catastrophic errors.

We have investigated in detail the quality of photometric redshifts as
a function of spectral type, apparent magnitude and redshift based on
the comparison with the VVDS spectroscopic redshifts. This step is
crucial for forthcoming scientific analysis. As expected we find that
the apparent magnitude is the key parameter: the fraction of
catastrophic errors increases by a factor 12 and the rms by a factor
1.7 between $17.5\le i'_{AB} \le 21.5$ and $23.5 \le i'_{AB} \le
24$. The reliability of the photometric redshifts also depends on the
spectral type: half of the catastrophic errors are galaxies which are
best fitted by a starburst template type. The evolution of $\eta$ as a
function of redshift shows that the redshift range the most reliable
for forthcoming scientific analysis is $0.2<zp<1.5$ for the complete
population of galaxies. This range can be extended into the ``redshift
desert'' when near-infrared data are available (although currently
only 6\% of the field is covered).

We present $i'$ band selected redshift distributions at $i'_{AB}\le 23$
and $i'_{AB}\le 24$ which are fully consistent with the redshift
distributions derived from the VVDS spectroscopic redshifts. We show
the ability of our method to correctly recover the redshift
distributions, including even the identification of the strongest
density peaks. We show that a near-infrared selected sample is very
efficient for the selection of high redshift galaxies, with 40\% of the
sample at $z>1$ for $K\le 23$.  This robust $K$ selected sample will be
used to investigate the evolution of the stellar mass function
(\cite{Pozzetti06}2006, in preparation) which is a crucial test of the
hierarchical model (e.g.  \cite{Kauffmann98}1998, \cite{Cimatti02}2002,
\cite{Somerville04}2004) of galaxy formation.

Finally, we have applied our robust photometric redshifts measurement
code on all four CFHTLS deep fields ({\it
  http://www.cfht.hawaii.edu/Science/CFHTLS}). We measure photometric
redshifts for an uniquely large and deep sample of 522286 objects at
$i'_{AB}\le 25$ on 3.2 deg$^2$. We assess the accuracy of these
photometric redshifts beyond the spectroscopic limits and we present
the redshift distributions in these four deep fields showing that
cosmic variance effects are present at the 15\% level for fields of
size 0.8 deg$^2$.

All photometric redshifts and input photometric catalogues are made
publicly available. 

\begin{acknowledgements}
  This research has been developed within the framework of the VVDS
  consortium.  This work has been partially supported by the CNRS-INSU
  and its Programme National de Cosmologie (France) and Programme
  National Galaxies (France), and by Italian Ministry (MIUR) grants
  COFIN2000 (MM02037133) and COFIN2003 (num.2003020150).  The VLT-VIMOS
  observations have been carried out on guaranteed time (GTO) allocated
  by the European Southern Observatory (ESO) to the VIRMOS consortium,
  under a contractual agreement between the Centre National de la
  Recherche Scientifique of France, heading a consortium of French and
  Italian institutes, and ESO, to design, manufacture and test the
  VIMOS instrument.
\end{acknowledgements}


\begin{thebibliography}{}


\bibitem[{Arnouts et al. }]{Arnouts99} Arnouts S., Cristiani S., Moscardini L. et al., 1999, MNRAS, 310, 540 
%\bibitem[{Arnouts et al. }{2001}]{Arnouts01} Arnouts S., Vandame B., Benoist C. et al., 2001, A\&A, 379, 740 
\bibitem[{Arnouts et al. }]{Arnouts02} Arnouts S., Moscardini L., Vanzella E. et al., 2002, MNRAS, 329, 355 

\bibitem[{Baum }]{Baum62}Baum W. A., 1962, IAU Symp., 15, 390

\bibitem[{Ben{\'{\i}}tez }]{Benitez00} Ben{\'{\i}}tez N., 2000, ApJ, 536, 571 
\bibitem[{Ben{\'{\i}}tez }]{Benitez04} Ben{\'{\i}}tez N., Ford H., Bouwens R. et al., 2004, ApJS, 150, 1 
 
\bibitem[{Bertin \& Arnouts }]{Bertin96} Bertin E.~\& Arnouts S., 1996, A\&AS, 117, 393 

\bibitem[{Bolzonella et al. }]{Bolzonella00} Bolzonella M., Miralles J.-M. \& Pell{\' o} R., 2000, A\&A, 363, 476 
\bibitem[{Bolzonella et al. }]{Bolzonella02} Bolzonella M., Pell{\' o} R., \& Maccagni D., 2002, A\&A, 395, 443 

\bibitem[{Brodwin et al. }]{Brodwin06} Brodwin M., Lilly S.J., Porciani C. et al., 2006, ApJS, 162, 20 

\bibitem[{Boulade et al. }]{Boulade03} Boulade O., Charlot X., Abbon P. et al., 2003, SPIE, 4841, 72 

\bibitem[{Bruzual et Charlot }]{Bruzual03} Bruzual G.~ et Charlot S., 2003, MNRAS, 344, 1000

\bibitem[{Budav{\' a}ri et al. }]{Budavari00} Budav{\' a}ri T., Szalay A.~S., Connolly A.~J., Csabai I. \& Dickinson M., 2000, AJ, 120, 1588 

\bibitem[{Coleman, Wu \& Weedman }]{Coleman80} Coleman G.D., Wu C.-C., Weedman D.W., 1980, ApJS, 43, 393 

\bibitem[{Connolly et al. }]{Connolly95} Connolly A.~J., Csabai I., Szalay A.~S., Koo D.~C., Kron R.~G., \& Munn J.~A., 1995, AJ, 110, 2655 
 
\bibitem[{Csabai et al. }]{Csabai00} Csabai I., Connolly A.~J., Szalay A.~S. \& Budav{\' a}ri T., 2000, AJ, 119, 69 

\bibitem[{Cimatti et al. }]{Cimatti02} Cimatti A., Pozzetti L., Mignoli M. et al., 2002, A\&A, 391, L1 

\bibitem[{Fern{\' a}ndez-Soto et al. }]{Fernandez99} Fern{\' a}ndez-Soto A., Lanzetta K.~M. \& Yahil A., 1999, ApJ, 513, 34 
%\bibitem[{Fern{\' a}ndez-Soto et al. }]{Fernandez01} Fern{\' a}ndez-Soto A., Lanzetta K.~M., Chen H., Pascarelle S.~M., \& Yahata N., 2001, ApJS, 135, 41 
%\bibitem[{Fern{\' a}ndez-Soto et al. }]{Fernandez02} Fern{\' a}ndez-Soto A., Lanzetta K.~M., Chen H.-W., Levine B. \& Yahata N., 2002, MNRAS, 330, 889 

\bibitem[{Fioc \& Rocca-Volmerange }]{Fioc97} Fioc M.~\& Rocca-Volmerange B., 1997, A\&A, 326, 950 

\bibitem[{Fontana et al. }]{Fontana00} Fontana A., D'Odorico S., Poli F. et al., 2000, AJ, 120, 2206 

\bibitem[{Gabasch et al. }]{Gabasch04} Gabasch A., Bender R., Seitzer S. al., 2004, A\&A, 421, 41 
 
%\bibitem[{Gwyn }]{Gwyn96} Gwyn S.~D.~J. \& Hartwick F.~D.~A., 1996, ApJL, 468, L77 

\bibitem[{Ilbert et al. }]{Ilbert05} Ilbert O., Tresse L., Zucca E. et al., 2005, A\&A, 439, 863

\bibitem[{Iovino et al. }]{Iovino05} Iovino A., McCracken H.J., Garilli B. et al., 2005, A\&A, in press, astro-ph/0507668
 
\bibitem[{James \& Roos } ]{James95} James F. \& Roos M., 1995, MINUIT Function Minimization and Error Analysis, Version 95.03, CERN Program Library D506

\bibitem[{Kauffmann \& Charlot }]{Kauffmann98} Kauffmann G. \& Charlot S., 1998, MNRAS, 297, L23 

\bibitem[{Kinney et al. }]{Kinney96} Kinney A.L., Calzetti D., Bohlin R.C., McQuade K., Storchi-Bergmann T. \& Schmitt H.R., 1996, ApJ, 467, 38 

\bibitem[{Kron }]{Kron80} Kron R.G., 1980, ApJS, 43, 305

\bibitem[{Le F\`evre et al. }]{LeFevre03} Le F{\`e}vre O., Saisse M., Mancini D. et al., 2003, SPIE, 4841, 1670
\bibitem[{Le F\`evre et al. }]{LeFevre04a} Le F{\` e}vre O.,  Mellier Y., McCracken H. J. et al., 2004a, A\&A, 417, 839 
\bibitem[{Le F\`evre et al. }]{LeFevre04b} Le F{\` e}vre O.,  Vettolani G., Paltani S. et al., 2004b, A\&A, 428, 1043 
\bibitem[{Le F\`evre et al. }]{LeFevre05a} Le F\`evre O., Vettolani G., Garilli B. et al., 2005a, A\&A, 439, 845
\bibitem[{Le F\`evre et al. }]{LeFevre05b} Le F\`evre O., Paltani P., Arnouts S. et al., 2005b, nature, 437, 519

\bibitem[{Madau }]{Madau95} Madau P., 1995, ApJ, 441, 18 

\bibitem[{McCracken et al. }]{McCracken03} McCracken H.J., Radovich M., Bertin E. et al., 2003, A\&A, 410, 17 
\bibitem[{McCracken et al. }]{McCracken06} McCracken H.~J.~et al., 2006, A\&A, in preparation

%\bibitem[{Menci et al. }]{Menci02} Menci N., Cavaliere A., Fontana A., Giallongo E. \& Poli F., 2002, ApJ, 575, 18 
%\bibitem[{Menci et al. }]{Menci04} Menci N., Cavaliere A., Fontana A., Giallongo E., Poli F. \& Vittorini V.\ 2004, ApJ, 604, 12 

\bibitem[{Mobasher et al. }]{Mobasher04} Mobasher B., Idzi R., Benítez N. et al., 2004, ApJL, 600, L167 


\bibitem[{Pickles }]{Pickles98} Pickles A.~J., 1998, PASP, 110, 863 

\bibitem[{Phillips et al. }]{Phillips97} Phillips A.C., Andrew C., Guzman R. et al., 1997, ApJ, 489, 543 

\bibitem[{Pozzetti et al. }]{Pozzetti06} Pozzetti L. et al., 2006, A\&A, in preparation
 
\bibitem[{Prevot et al. }]{Prevot84} Prevot M.L., Lequeux J., Prevot L., Maurice E. \& Rocca-Volmerange B., 1984, A\&A, 132, 389 
 
\bibitem[{Puschell }]{Puschell82} Puschell J.J., Owen F.N. \& Laing R.A., 1982, ApJL, 257, L57 

\bibitem[{Sawicki et al. }]{Sawicki97} Sawicki M.J., Lin H. \& Yee H.K.C., 1997, AJ, 113, 1 

\bibitem[{Schlegel et al. }]{Schlegel98} Schlegel D.~J., Finkbeiner D.~P., \& Davis M., 1998, ApJ, 500, 525 
 
\bibitem[{Somerville et al. }]{Somerville04} Somerville R.S., Moustakas L.A., Mobasher B. et al., 2004, ApJL, 600, L135 
 
\bibitem[{Vanzella et al. }]{Vanzella04} Vanzella E., Cristiani S., Fontana A. et al., 2004, A\&A, 423, 761 

\bibitem[{Wang et al. }]{Wang98} Wang Y., Bahcall N. \& Turner E.L., 1998, AJ, 116, 2081 
 
\bibitem[{Wolf et al. }]{Wolf03} Wolf C., Meisenheimer K., Rix H.-W. et al., 2003, A\&A, 401, 73 

\bibitem[{Wolf et al. }]{Wolf04} Wolf C., Meisenheimer K., Kleinheinrich M. et al., 2004, A\&A, 421, 913 


\end{thebibliography}
\end{document}